\documentclass[aps,twocolumn,pra,superscriptaddress,letterpaper,nofootinbib]{revtex4-1}
\usepackage{amssymb}
\usepackage{physics}
\usepackage{mathtools}
\usepackage{txfonts}
\usepackage[pdftex]{graphicx}
\usepackage[usenames, dvipsnames, svgnames, table]{xcolor}
\usepackage[colorlinks=true, citecolor=Blue,linkcolor=BrickRed, urlcolor=magenta]{hyperref}
\usepackage{bm}

\begin{document}


\newcommand{\R}[1]{\textcolor{red}{#1}}
\newcommand{\B}[1]{\textcolor{blue}{#1}}
\newcommand{\finesse}{\mathcal{F}}
\def\*#1{\bm{#1}}
\def\fw#1{\texttt{#1}}


\title{Fundamental limitations of cavity-assisted atom interferometry}

\author{M. Dovale-\'Alvarez}
\email{mdovale@star.sr.bham.ac.uk}
\author{D. D. Brown}
\author{A. W. Jones}
\author{C. M. Mow-Lowry}
\author{H. Miao}
\author{A. Freise}

\affiliation{School of Physics and Astronomy and Institute of Gravitational Wave Astronomy, University of Birmingham, Edgbaston, Birmingham B15 2TT, United Kingdom}


\begin{abstract}

Atom interferometers employing optical cavities to enhance the beam splitter pulses promise significant advances in science and technology, notably for future gravitational wave detectors. Long cavities, on the scale of hundreds of meters, have been proposed in experiments aiming to observe gravitational waves with frequencies below 1 Hz, where laser interferometers, such as LIGO, have poor sensitivity. Alternatively, short cavities have also been proposed for enhancing the sensitivity of more portable atom interferometers. We explore the fundamental limitations of two-mirror cavities for atomic beam splitting, and establish upper bounds on the temperature of the atomic ensemble as a function of cavity length and three design parameters: the cavity $g$ factor, the bandwidth, and the optical suppression factor of the first and second order spatial modes. A lower bound to the cavity bandwidth is found which avoids elongation of the interaction time and maximizes power enhancement. An upper limit to cavity length is found for symmetric two-mirror cavities, restricting the practicality of long baseline detectors. For shorter cavities, an upper limit on the beam size was derived from the geometrical stability of the cavity. These findings aim to aid the design of current and future cavity-assisted atom interferometers.

\end{abstract}

\maketitle

\section{Introduction}

Since the demonstration of the first light-pulse atom interferometer~\cite{Kasevich1991}, this field has received great interest and has found applications across many areas of science and technology. It has made possible new measurements of the gravitational constant~\cite{Fixler2007, Rosi2014} and the fine structure constant~\cite{Weiss1994,Cadoret2008, Bouchendira2011}, and tests of the weak equivalence principle~\cite{Schlippert2014, Tarallo2014}. It has seen an important development for applications in inertial sensing, to measure gravity accelerations~\cite{Peters1999,Hu2013}, gravity gradients~\cite{McGuirk2002, Sorrentino2014}, and rotations~\cite{Gustavson1997, Canuel2006, Stockton2011, Dutta2016}. It also has proposed applications in tests of general relativity~\cite{Dimopoulos2007, Dimopoulos2008}, quantum electrodynamics~\cite{Wolf2007}, and quantum-entanglement at macroscopic distances~\cite{Kovachy2015a}. Perhaps the most tantalizing application of all is in gravitational wave astronomy, where atom interferometers are expected to observe gravitational waves with frequencies below 1\,Hz~\cite{Dimopoulos2008a,Graham2013,Chaibi2016,Canuel2016,Canuel2017}, a frequency band forbidden in the most advanced optical interferometers, such as Advanced LIGO~\cite{Collaboration2015}.

In light pulse atom interferometry, atomic beams are coherently split and later recombined using laser pulses as beam splitters. The sensitivity of these devices increases with the measured phase difference between the matter waves, which scales with the relative momentum between the two arms of the interferometer and the free evolution time between pulses. 
In large momentum transfer (LMT) interferometry the atoms coherently scatter $2n$ photons from the laser beams and acquire a momentum difference of $2n \hbar k$. However, the increased number of photon-atom interactions means that the sensitivity to inhomogeneities of the relative laser phase is $n$ times higher than that of a conventional interferometer.
LMT methods include sequential Raman pulses~\cite{McGuirk2000}, sequential two-photon Bragg diffraction~\cite{Chiow2011}, and multi-photon Bragg diffraction~\cite{Mueller2008}. The latter has the advantage of achieving large momentum transfer using a single laser pulse while leaving the internal energy state of the atom unchanged, leading to the cancellation of important systematic effects. In addition to the increased sensitivity to the relative laser phase, this method is limited by the available laser power.

\begin{figure}
\centering
\includegraphics[scale=0.1275]{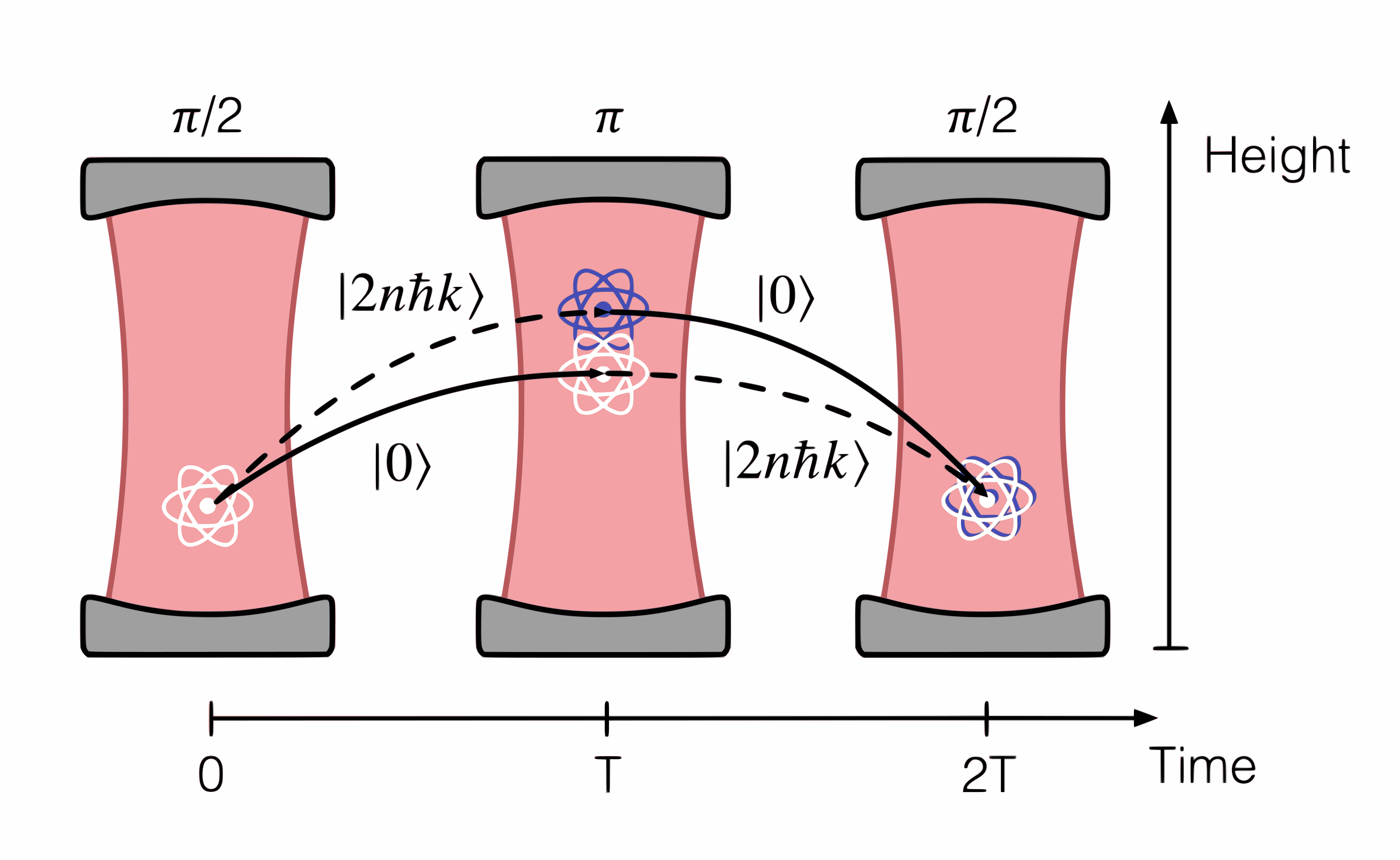}
\caption{An interferometer sequence using cavity-assisted large momentum beam splitters. A cloud of atoms is coherently split ($t=0$), deflected ($t=T$), and recombined ($t=2T$) using light pulses that resonate in the cavity.}
\label{figure:cavity_setup}
\end{figure}

Optical cavities are proposed as the key enabling technology for LMT beam splitters, as performing the interferometric sequence inside the cavity (Fig.~\ref{figure:cavity_setup}) can help mitigate the disadvantages of the technique: cavities provide spatial filtering of the interferometry beam, thus ``cleaning'' the optical wavefronts, and resonant enhancement in the cavity means that a high intracavity power may be achieved using a relatively low input power. 

Intracavity atom interferometry was demonstrated in~\cite{Hamilton2015}, where they show a $\pi/2-\pi-\pi/2$ interferometer with cesium atoms loaded horizontally into a vertical 40\,cm cavity. In this proof of principle experiment, the small cavity mode volume placed a tight constraint on the total measurement time, which was just $20\,$ms. The same group was able to increase the total measurement time up to 130\,ms~\cite{Hamilton2015a, Jaffe2017, Haslinger2017}. A clever design of a marginally stable cavity with an intracavity lens was proposed in~\cite{Riou2017}, also employing a perpendicular loading scheme but with a large mode volume capable of accommodating a 1\,$\mu$K cloud as it expands for up to 250\,ms. Cavity-assisted LMT beam splitters are also proposed for the gravitational wave antenna MIGA~\cite{Canuel2016, Canuel2017}, where the interferometric pulses resonate inside two horizontal 200\,m cavities and interrogate three atom clouds launched vertically for a total measurement time of 250\,ms.

Despite its promising nature, the advantages and limitations of cavity-assisted atom interferometry have not yet been quantified. The cavity bandwidth plays a major role in the performance of the interferometer. Power enhancement and spatial filtering are both enhanced by increasing the cavity finesse. The maximal allowed beam size increases with cavity length. In atom interferometry, both good spatial filtering and large beam sizes are desired qualities. The cavity bandwidth scales inversely with the product of finesse and length. Thus, it would seem obvious that the narrower the bandwidth is, the better. We find, however, that there is a limit to the bandwidth below which the pulses suffer severe elongation ---leading to undesirably long interaction times--- and power enhancement of the beam splitter pulses worsens dramatically, nullifying the advantage of incorporating the cavity in the first place. Having realized this bandwidth limit, the task then becomes a balancing act between the quality of the cavity as a spatial filter of the interferometric beams and its ability to accommodate the size of the atomic cloud as it thermally expands during the measurement.

This paper is structured as follows: In Sec.~II we introduce the atom optics model of the cavity. In Sec.~III we present the simulation results, explaining the effect that the cavity parameters have on the atomic transitions. In Sec.~IV we treat the problem from a purely geometrical and optical perspective, analyzing the quality of the cavity as a spatial filter of the interferometric beams.

\section{Atom Optics Model of the Cavity}

\begin{figure}
\centering
\includegraphics{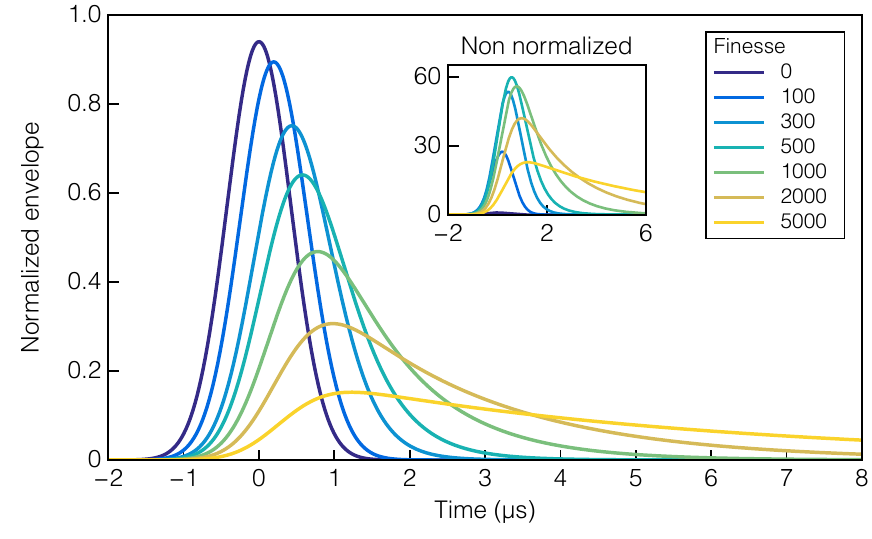}
\caption{Cavity-induced deformation of a Gaussian input. Envelope functions of the intracavity field for a 1\,m cavity injected with a 1\,$\mu$s pulse for different cavity finesses. All areas are normalized to the input pulse area for comparison. When the pulse duration is comparable to the photon lifetime of the cavity, its envelope function is elongated. \emph{Inset}: Envelopes without normalization. }
\label{figure:cavity_envelope_deformation}
\end{figure}

We build a computer model to simulate the outcome of the photon-atom interactions inside the optical cavity. In order to keep the effect of velocity selectivity~\cite{Moler1992} to a minimum, the duration of the beam splitter and mirror pulses is set as short as possible. The interaction time must be long enough to obtain efficient population transfer to the excited state with low losses, but as short as possible to maximize atomic flux through the pulse sequence. This interaction regime is known as the quasi-Bragg regime~\cite{Mueller2008a}, and unfortunately there is no shortcut for solving the equations of motion of the amplitudes of the atomic states,
\begin{equation}
	i \dot g_n = \left[\omega_r n^2 + \Omega \right] g_n + \frac{1}{2} \Omega \left[g_{n+2}+g_{n-2} \right],
\label{equation:Bragg}
\end{equation}
where $g_n$ is the amplitude of the state with momentum $n \hbar k$, $\omega_r$ is the recoil frequency, and $\Omega$ is the two-photon Rabi frequency, which is in general an arbitrary function of time. For a pulsed field we generally write $\Omega(t) = \bar \Omega G(t)$, where $\bar \Omega$ is the peak two-photon Rabi frequency ---proportional to the peak laser intensity--- and $G(t)$ is the envelope function of the field intensity, of full width at half maximum (FWHM) $\delta t$. Hereafter we will simply refer to the FWHM of the pulse as pulse width or duration.

In this interaction regime the shape of $G(t)$ plays a major role in the evolution of the atomic states. For example, square pulses are known to give high losses of the population into the intermediate states~\cite{Keller1999}, whereas pulses having smooth envelope functions like the Gaussian can produce efficient transitions even at short interaction times. 

The photon lifetime of the cavity is defined as the time it takes for the circulating intensity to drop by a factor $1/e$ after the input is abruptly switched off. For an impedance matched and lossless cavity it is approximated by $\tau_c = L \finesse / \pi c$, where $L$ is the length of the cavity, $\finesse$ is the finesse, and $c$ is the speed of light. If the pulse duration is comparable to the photon lifetime of the cavity, the circulating field will present a deformed $G(t)$, asymmetric and with a larger area and width than the input pulse (Fig.~\ref{figure:cavity_envelope_deformation}). For example, for a cavity with a photon lifetime of $1\,\mu$s injected with a short pulse of width $\delta t = 1\,\mu$s, the intracavity field presents a deformed envelope of width $\approx 1.8\,\mu$s, 80\% longer than the input. A complete account of the photon-atom interactions in the optical cavity must include this effect.

\begin{figure}
\centering
\includegraphics{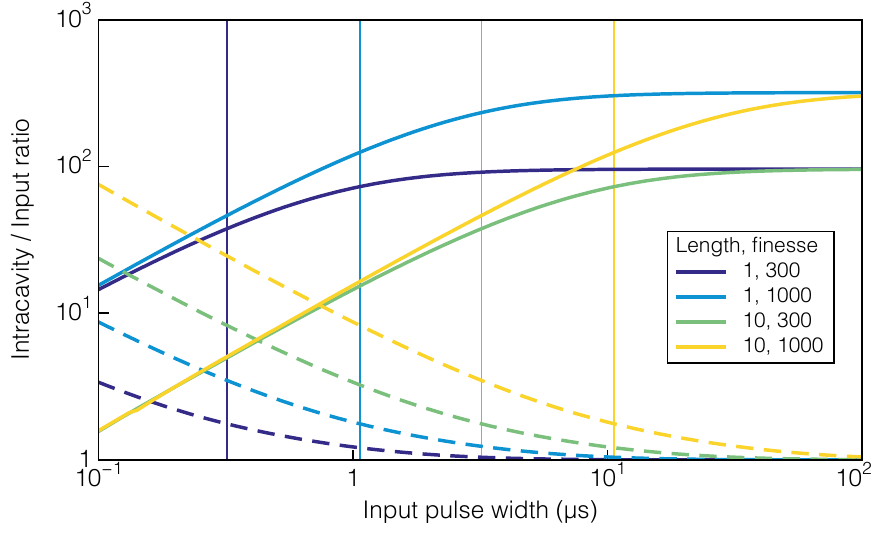}
\caption{Transient response of the cavity to short pulses. Intracavity to input ratios of the pulse area (solid lines) and width (dashed lines) vs.\ input pulse width, for four cavities of different length and finesse. The vertical lines represent the photon lifetime of each cavity.}
\label{figure:cavity_ratios}
\end{figure}

Cavities having different photon lifetimes --- or, equivalently, different bandwidth --- respond differently to the same input (Fig.~\ref{figure:cavity_ratios}). For input pulse widths much larger than the photon lifetime of the cavity, $\delta t \gg \tau_c$, the circulating field envelope tends to match the input envelope multiplied by the optical gain. In this scenario the cavity offers maximum power enhancement and does not cause any elongation, i.e., the temporal profile of the intracavity field is dominated by the input. For pulse durations on the order of the photon lifetime or lower, power enhancement drops as the cavity does not reach a steady state, and the circulating field becomes elongated, with a temporal profile dominated by the cavity. As the input width approaches $\tau_c$, the elongation becomes more severe and the circulating power tends to match the input power.

\begin{figure}
\centering
\includegraphics{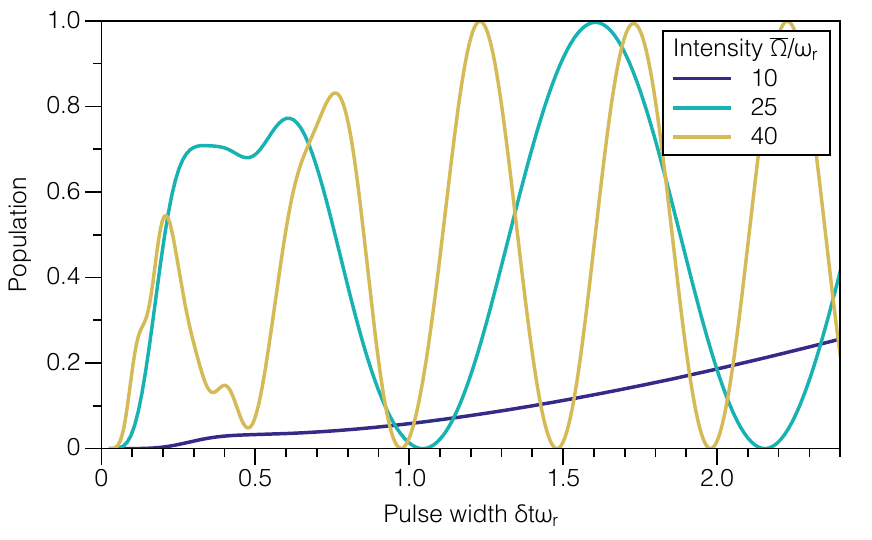}
\caption{Population of state $\ket{+4 \hbar k}$ as a function of pulse width for a Gaussian envelope, for different values of the laser intensity. When the interaction time is short ($\delta t \omega_r \lessapprox 1$) the intermediate states cannot be adiabatically eliminated and the equations of motion need to be solved numerically considering all intermediate states and sufficiently many outer states.}
\label{figure:transitions}
\end{figure}

Our model solves Eq.~(\ref{equation:Bragg}) for a truncated set of states $\ket{-n-m}...\ket{+n+m}$, where $n$ is the order of the $2n$-photon Bragg diffraction process and $m$ is the number of additional outer states considered. We find that considering four outer states is usually sufficient for $n \leq 10$, i.e., the solution is not altered by considering more. Since Eq.~(\ref{equation:Bragg}) couples even and odd states separately, we look only at solutions with either all even or odd terms zero. The model is checked against the known analytical solutions for first order Bragg diffraction and Bragg diffraction in the Raman-Nath regime, and it also reproduces the results presented in~\cite{Riou2017}. See Fig.~\ref{figure:transitions} for an example of a $n=4$ process in which we scan the pulse width at fixed intensity and plot the population of the final state, $\ket{+4 \hbar k}$. Throughout this article, we present all results in terms of the dimensionless interaction time or pulse width $\delta t \omega_r$, and the dimensionless interaction strength or intensity $\bar \Omega / \omega_r$. This makes all results readily scalable for the atomic transition of interest, with $\omega_r = \hbar k^2 / 2M$, where $M$ is the mass of the atom. For example, $\omega_r = 23694\,$Hz for the rubidium-87 D$_2$ transition ($5^2$S$_{1/2}\rightarrow 5^2$P$_{3/2}$), and $\omega_r = 12983\,$Hz for the cesium-133 D$_2$ transition ($6^2$S$_{1/2}\rightarrow 6^2$P$_{3/2}$).

\section{Effect of the Cavity on the Atomic Transitions and Cavity Bandwidth Limit}

\begin{figure}[b!]
\centering
\includegraphics{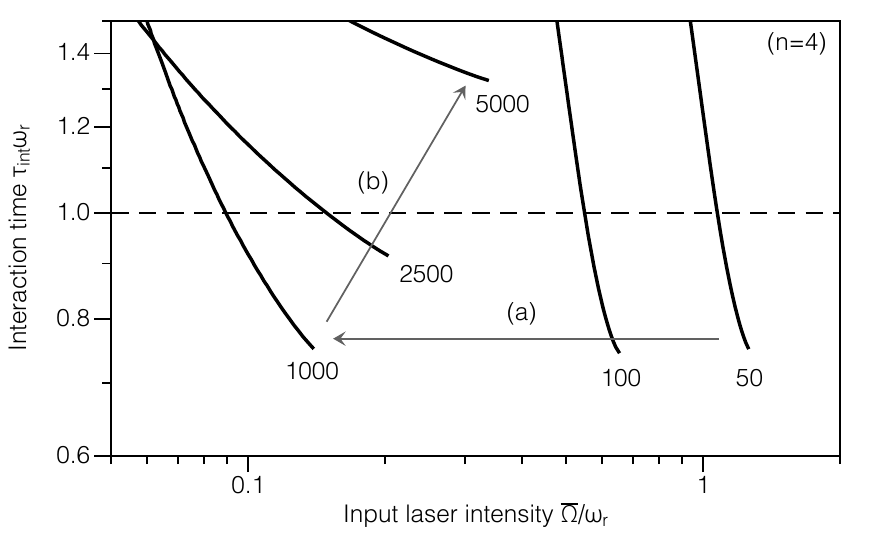}
\caption{Duration vs intensity of the first mirror pulses in the 10\,m cavity for different values of the cavity finesse, indicated at the bottom of each curve. As the cavity finesse increases, the curves shift left as the beam splitters require less input power due to the increased optical gain (a). After reaching a particular value of the finesse, $\finesse_{\rm{max}}$, the curves shift right and up, as the cavity-induced elongation becomes more severe and power enhancement of the beam splitters worsens (b).}
\label{figure:cavity_finesse_effect}
\end{figure}

\begin{figure}
\centering
\includegraphics{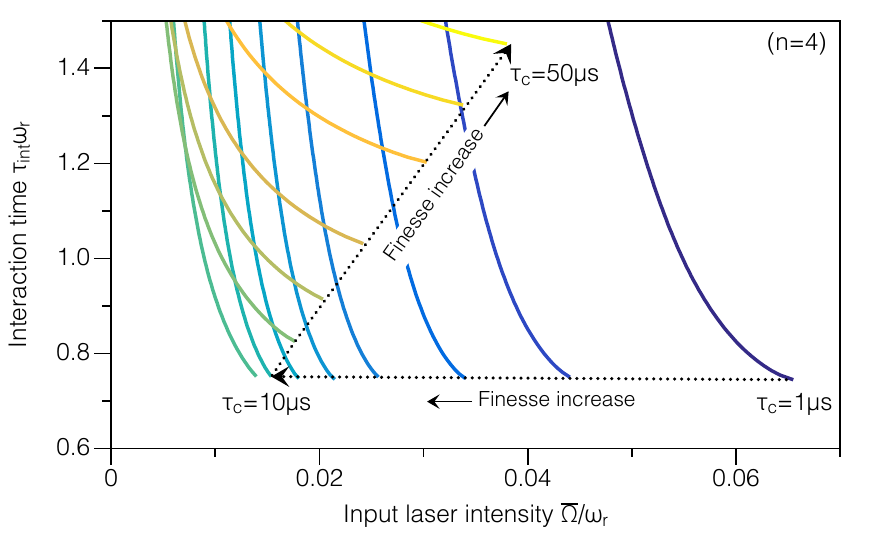}
\caption{Duration vs intensity of the first mirror pulses for varying cavity photon lifetimes. The required input power is minimized for $\tau_c = \tau_{\rm{max}}$, in this case ($n=4$) $\tau_{\rm{max}} \approx 10\,\mu$s. For $\tau_c > \tau_{\rm{max}}$, the minimum interaction time grows significantly. The input laser intensity scale shown here can be adjusted for any cavity length $L$ by applying a factor $L/1\,$m.}
\label{figure:n=4}
\end{figure}

Without loss of generality, we consider the interaction of the circulating field in a 10\,m cavity, injected with a pulse of Gaussian envelope function, with an ensemble of cold rubidium-87 atoms. For given cavity parameters, the population transfer efficiency to the target state, $\mathcal P$, is a function of the input pulse's intensity and width:
\begin{equation}
	\mathop {\lim}\limits_{t \to +\infty} \left| \bra{+n}\ket{\psi } \right|^2 = \mathcal P(\bar \Omega, \delta t).
\end{equation}
where $\ket{\psi} = (...g_{-n},g_{-n+2}...)$ is the wavefunction describing the state of the atom. We determine the mirror pulse durations, $\delta t_{\pi}$, to transfer the ensemble to the target state with losses below 5\%: 
\begin{equation}
	\delta t_{\pi}(\bar \Omega) = \min \left\{ \delta t : \mathcal P(\bar \Omega, \delta t)  > 0.95 \right\}
\end{equation}
We restrict our computation to the first Rabi cycle ---hence the `$\min$'--- for simplicity. This is a reasonable restriction, as the interferometer requires the interaction time to be as short as possible. We perform this computation as we change the finesse of the 10\,m cavity and measure the width of the elongated intracavity pulse $\tau_{\pi}$ (Fig.~\ref{figure:cavity_finesse_effect}). Note that $\tau_{\pi}$ is the actual interaction time, and not $\delta t_{\pi}$ which refers to the width of the injected excitation.

As we increase the cavity finesse we note three effects: (1) The required laser intensity of the beam splitters becomes considerably lower, as expected due to the cavity's buildup effect and highlighting the advantage of cavities for LMT beam splitting. (2) Power enhancement reaches a maximum for some value of the finesse, $\finesse_{\rm{max}}$; increasing the finesse further comes at the price of increased intensity requirements for the short pulses. (3) The duration of the mirror pulses stays roughly the same as if there was no cavity up to $\finesse_{\rm{max}}$; increasing the finesse further also comes at the price of increased interaction times, as the cavity's elongation effect becomes more severe.

The cavity's effect on the pulse is therefore projected onto the transition probabilities by shifting both the required laser intensity and the photon-atom interaction time. Moreover, simulating more cavity length and finesse ranges and additional diffraction orders, we find that the shape of $\mathcal P(\bar \Omega, \delta t)$ is a function of the cavity bandwidth only. I.e., there is no distinction between a length change and a finesse change with the exception of a linear shift in $\bar \Omega$ due to the scaling in optical gain. This is expected, as the cavity bandwidth univocally determines the shape of $G(t)$. Therefore, $\finesse_{\rm{max}}$ can be extrapolated for any cavity length from the value of the cavity photon lifetime $\tau_{\rm{max}}$. For example, cavities with length $L$ will exhibit the same behavior depicted in Fig.~\ref{figure:cavity_finesse_effect} for finesses adjusted by the ratio $10\,$m/$L$ and intensities adjusted by $L/10\,$m.

In the absence of the cavity, the values of $\delta t_{\pi}(\bar \Omega)$ decrease slightly with increasing $n$ for $n>2$, i.e., higher order processes yield shorter beam splitter pulses, which in turn have greater intensity requirements. The cavity deforms the pulse's envelope function $G(t)$, as determined solely by $\tau_c$, and the cavity with $\tau_c=\tau_{\rm{max}}$ presents an optimal $G(t)$ that minimizes the required input power of the beam splitters. The value $\tau_{\rm{max}}$ is observed, through simulation, to be approximately 1/3 the duration of the shortest beam splitter pulse in the absence of the cavity. We believe this is because the optimal $G(t)$ occurs at a certain ratio between the cavity photon lifetime and the input pulse width, before the pulse gets significantly distorted by the cavity. Hence, the dependence of $\tau_{\rm{max}}$ on $n$ is roughly the same as that of $\min\left\{\delta t_{\pi}(\bar \Omega)\right\}/3$, which is a rather slow dependence.

\begin{figure}
\centering
\includegraphics{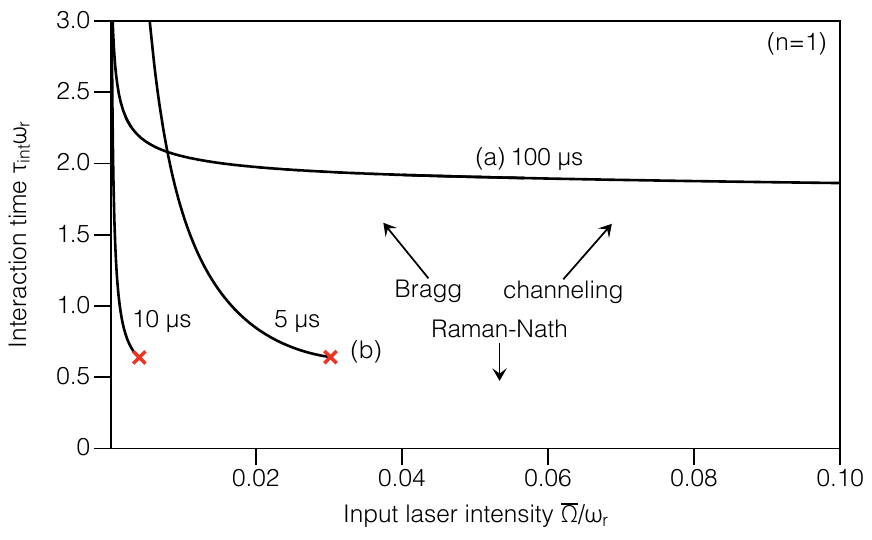}
\caption{First Rabi cycle of a conventional $2\hbar k$ beam splitter for three cavities with different photon lifetimes, indicated next to each curve. This transition is less lossy because of the direct coupling between the initial and final states. High-finesse cavities, with increased interaction times due to the severe elongation effect, exhibit a sharp transition between the adiabatic Bragg regime and the long-interaction steep-potential channeling regime, embodying the uncertainty relation between time and energy in the parameter space (a). The crosses represent the points where the transfer efficiency falls below 95\% for the lower finesse cavities, as the interaction time approaches the Raman-Nath regime (b).}
\label{figure:n=1}
\end{figure}

\begin{figure}[b!]
\centering
\includegraphics{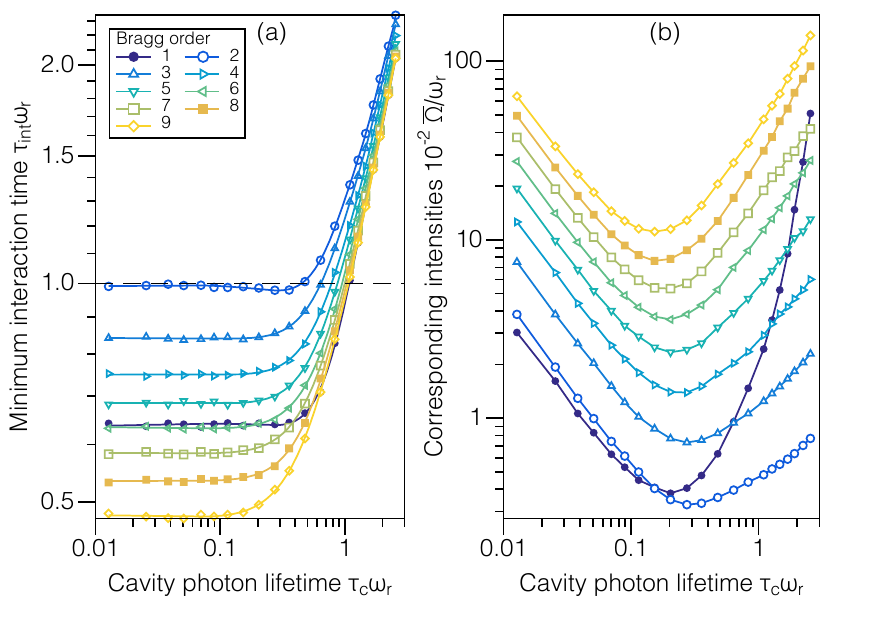}
\caption{Effect of the cavity photon lifetime $\tau_c$ on the atomic transitions. The minimum photon-atom interaction time remains largely unaffected for cavities with $\tau_c < \tau_{\rm{max}}$, and increases linearly for $\tau_c > \tau_{\rm{max}}$ (a). The required intensity for the shortest beam splitter is minimal for $\tau_c = \tau_{\rm{max}}$ (b). }
\label{figure:allorders}
\end{figure}

{\setlength{\tabcolsep}{0.25cm}
\begin{table*}
\small
\centering
\begin{tabular}{ ccccccccccccc }
\hline \hline
 & $\tau_{\rm{max}}\omega_r$ & $\tau_{\rm{max}}^{\rm{Rb87}}$ & $\tau_{\rm{max}}^{\rm{Cs133}}$ & $\Delta \omega_{\rm{min}}/\omega_r$ & $\Delta \omega_{\rm{min}}^{\rm{Rb87}}$ & $\Delta \omega_{\rm{min}}^{\rm{Cs133}}$ & \multicolumn{3}{c}{$\finesse_{\rm{max}}^{\rm{Rb87}}$} & \multicolumn{3}{c}{$\finesse_{\rm{max}}^{\rm{Cs133}}$} \\
$~n~$ &  & $(\mu\rm{s})$ & $(\mu\rm{s})$ &  & $(\rm{kHz})$ & $(\rm{kHz})$ & $1\,$m & $10\,$m & $100\,$m & $1\,$m & $10\,$m & $100\,$m \\ \hline
1 & 0.24 & 10.0 & 18.5 & 0.66 & 16.0 & 8.6 & 9373 & 934 & 91 & 17392 & 1736 & 171 \\
2 & 0.30 & 12.4 & 23.1 & 0.53 & 12.8 & 6.9 & 11717 & 1169 & 114 & 21741 & 2171 & 214 \\
3 & 0.28 & 11.6 & 21.5 & 0.56  & 13.7 & 7.4 & 10936 & 1091 & 106 & 20292 & 2026 & 200 \\
4 & 0.25 & 10.4 & 19.2 & 0.64 & 15.3 & 8.3 & 9763  & 974  & 94  & 18117 & 1809 & 178 \\
5 & 0.22 & 9.1 & 16.9  & 0.72 & 17.4 & 9.4 & 8592  & 856  & 83 & 15943 & 1591 & 156 \\
6 & 0.20 & 8.3 & 15.4  & 0.80 & 19.2 & 10.3 & 7810  & 778  & 75 & 14493 & 1446 & 142 \\
7 & 0.19 & 7.8 & 14.4  & 0.85 & 20.5 & 11.1 & 7302 & 727 & 70 & 13551 & 1352 & 132 \\
8 & 0.18 & 7.3 & 13.5  & 0.90 & 21.9 & 11.8 & 6834 & 681 & 65 & 12681 & 1265 & 124 \\
9 & 0.16 & 6.6 & 12.3  & 0.99 & 24.0 & 12.9 & 6248 & 622 & 59 & 11594 & 1157 & 113 \\
\hline \hline
\end{tabular}
\vspace{0.25cm}
\caption{Maximal cavity parameters for atom optics. A cavity with photon lifetime $\tau_{\rm{max}}$ (or bandwidth $\Delta \omega_{\rm{min}}$) minimizes the required power of the atomic beam splitters and keeps the interaction time unaffected. The corresponding finesse $\finesse_{\rm{max}}$ is given for cavity lengths of 1, 10 and 100 meters.}
\label{table:thetable}
\end{table*}

By increasing the interaction time, increasing the cavity finesse has the effect of parametrically pushing the photon-atom interactions towards the Bragg and channeling regimes. An evidence of this is the change in the slope of $\tau_{\pi} (\bar \Omega)$ for cavities with $\tau_c > \tau_{\rm{max}}$, as can be seen in Fig.~\ref{figure:n=4}. The very high finesse cavities have a slope $d \tau_{\pi} / d \bar \Omega \rightarrow +\infty$ for $\bar \Omega \rightarrow 0$, indicating adiabacity, and $d \tau_{\pi} / d \bar \Omega \rightarrow 0$ otherwise, indicating the channeling effect. As the cavity storage time becomes higher, the atomic interactions with the circulating cavity field become inevitably longer, and as they do so the diffraction process becomes more adiabatic. A high finesse cavity will transform a short input pulse with a large energy uncertainty into a long pulse with a well-defined energy. In doing so, energy conservation will favor transitions to the target state with low losses, unless the price is paid in terms of input power to drive efficient transitions that violate the adiabacity condition, thus operating in the long-interaction steep-potential channeling regime. 

\begin{figure}
\centering
\includegraphics{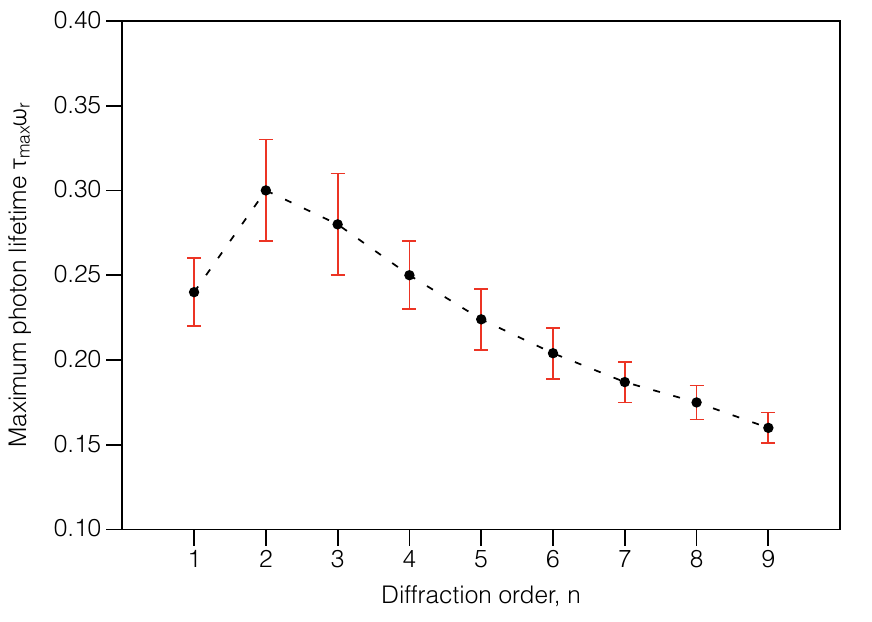}
\caption{Variation of $\tau_{\rm{max}}$ with the order of the diffraction process. The error bars represent the statistical uncertainty yielded by propagation of error through a least squares fit of the data to calculate the photon lifetimes that minimize the required input intensity of the beam splitters.}
\label{figure:tau}
\end{figure}

This is best seen in first order Bragg diffraction, $n=1$ (Fig.~\ref{figure:n=1}). This case is unique because the initial and final states, $\ket{\pm \hbar k}$, are coupled directly. While both states can transfer population to and from their other neighbor, the direct coupling between them makes the transition naturally less lossy. For this reason the elongation effect appears to be less significant in this case when compared to, e.g., $n=2$, but in turn the intensity required for the shortest beam splitters grows more steeply. The adiabacity condition is lower for $n=1$, i.e., the gap separating the Bragg and channeling regimes is narrower, and the higher finesse cavities with increased interaction times are parametrically ``pushed'' to transition sharply between the two regimes. In Fig.~\ref{figure:n=1} note how the cavity with $\tau_c = 100\,\mu$s produces efficient transitions with $d \tau_{\pi} / d \bar \Omega \rightarrow 0$ within the first Rabi cycle, whereas at lower $\tau_c$ efficiencies fall below 95\% as expected in the quasi-Bragg regime. The curve for $\tau_c = 100\,\mu$s in Fig.~\ref{figure:n=1} is a good representation of the well known uncertainty relation between time and energy in this parameter space.

To summarize, there is a value of the cavity bandwidth $\Delta \omega_{\rm{min}} = (2 \pi \tau_{\rm{max}})^{-1}$ which guarantees that the cavity offers maximum power enhancement of the short beam splitter pulses while keeping the interaction time unaffected (Fig.~\ref{figure:allorders}). A higher or lower bandwidth will incur in the requirement of a larger laser power, with lower bandwidths also incurring in longer interaction times. It is thus reasonable to treat $\Delta \omega_{\rm{min}}$ as a lower bound to cavity bandwidth or, equivalently, $\tau_{\rm{max}}$ as an upper bound to the cavity photon lifetime. Note that these bandwidth limits are specially significant for long-baseline experiments, as the larger cavity length vastly reduces the feasible finesse range. Of course, it may be a design choice to use a bandwidth lower than $\Delta \omega_{\rm{min}}$ and suffer the consequences (e.g., greater power requirements and a reduction of atomic flux due to the elongated pulses being able to resolve the velocity spread of the cloud). The bandwidth limit obtained here is not a physical limit, but a design limit based on optimizing the performance of cavity-assisted atomic beam splitters.

The values of $\tau_{\rm{max}}$ for $n=1-9$ are depicted in Fig.~\ref{figure:tau}. The values for rubidium-87 for $n=1$ are easy to remember and very similar to the $n=4$ case: $10\,\mu$s photon lifetime (16\,kHz bandwidth), which translates into a $\finesse_{\rm{max}}$ of roughly 10000, 1000 and 100 for cavity lengths of 1, 10 and 100\,m respectively. See Table~\ref{table:thetable} for a complete set of values for $n=1-9$.

\section{Geometrical and optical limits of the cavity-assisted atom interferometer}

The biggest constraint on the cavity parameters is set by the fact that it must accommodate the size of the atomic cloud as it thermally expands during the measurement. In this section we study the limitations of the cavity as a spatial filter under this constraint. The requirement of having a large waist may lead the cavity to be pushed very close to the edge of geometrical instability, which carries the consequent problems of increased sensitivity to alignment errors, mirror surface imperfections, and coupling to higher-order spatial modes. In addition, there is an incompatibility between having a cavity with a large beam size and simultaneously good spatial filtering (Fig.~\ref{figure:bandwidthlimit}). These findings, along with those from the previous section, allow us to establish upper bounds on the temperature of the atomic ensemble as a function of cavity length and three design parameters.

We assume here a cavity with a symmetric two-mirror configuration. Having the beam waist at the center of the cavity means that the curvature of the beam is symmetric with respect to it, allowing the possibility of running the interferometric sequence along the optical axis (\emph{on-axis} sequence). In this configuration the cavity can be used to simultaneously interrogate two clouds launched vertically in a juggling atomic fountain. The cavity can also be used to interrogate several atom interferometers running in parallel along the optical axis with the clouds being loaded perpendicularly into the cavity, as proposed for MIGA (\emph{perpendicular} sequence). In on-axis sequences the total measurement time scales with $\sqrt{L}$, as the atoms explore some fraction of the cavity length, whereas in perpendicular sequences the total measurement time is a parameter independent of cavity length. 

\begin{figure}
\centering
\includegraphics{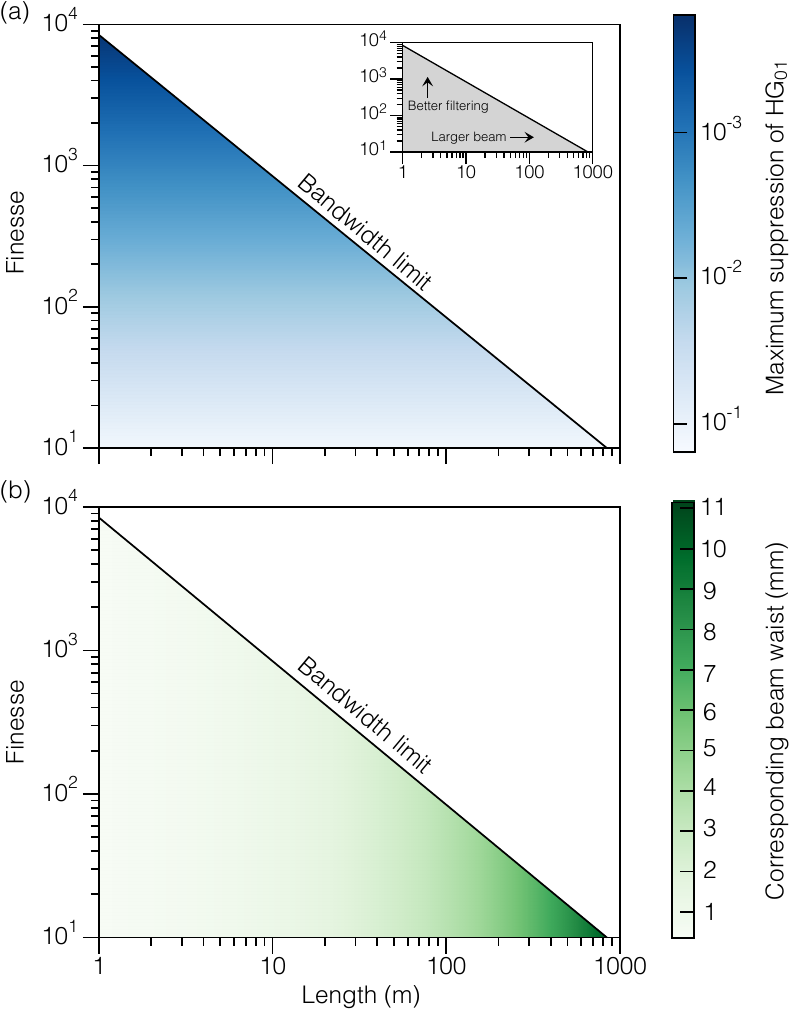}
\caption{The bandwidth limit places a constraint on the cavity's $L-\finesse$ parameter space, depicted here for $n=1$. Higher finesses lead to better spatial filtering (a), while longer lengths allow for larger beams (b).
 In atom interferometry, both large beams and good spatial filtering are desired qualities.}
\label{figure:bandwidthlimit}
\end{figure}

\begin{figure}
\centering
\includegraphics{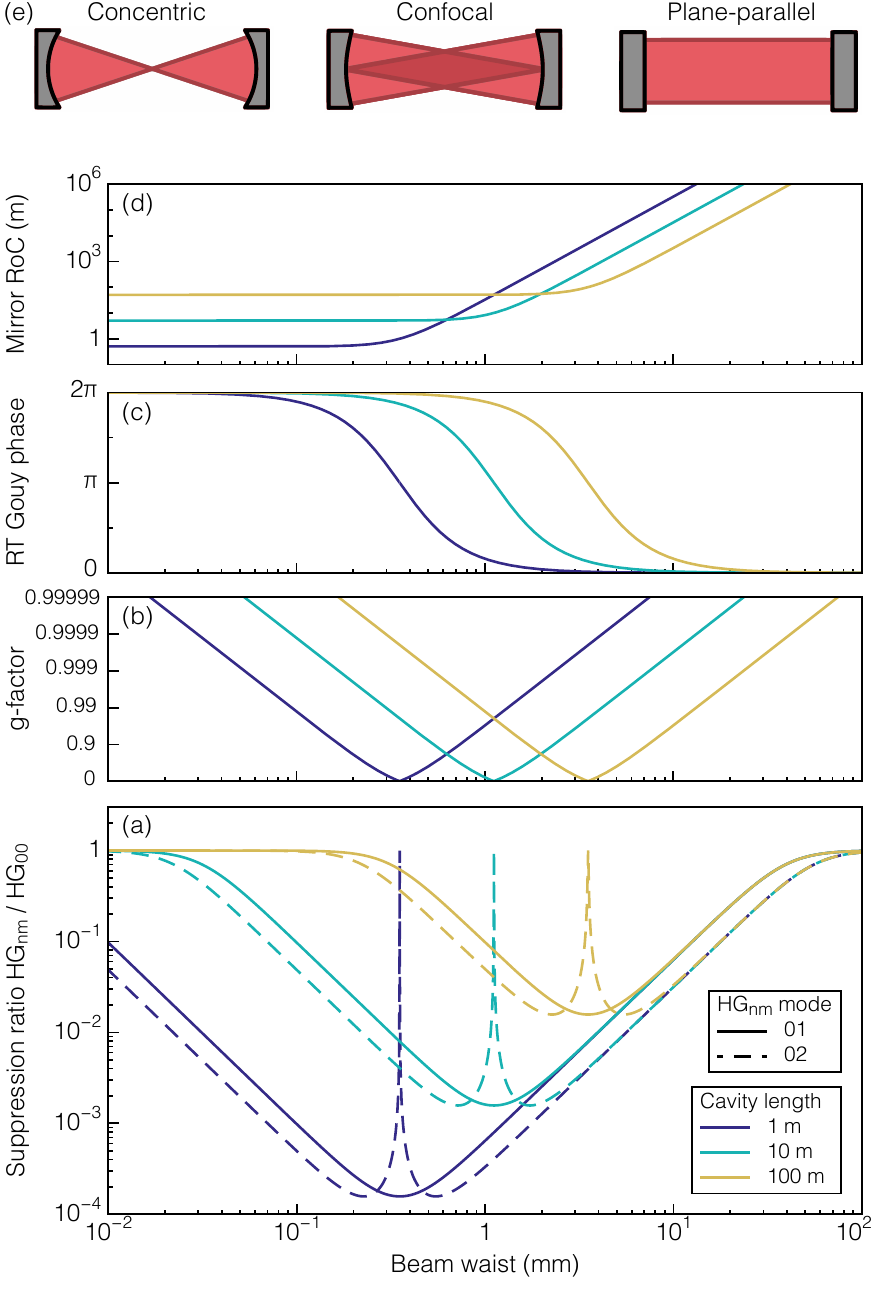}
\caption{Spatial filtering and geometrical properties of the cavity. The optical suppression factor of the first and second TEM modes serve as indication of the quality of the laser wavefronts as a function of beam waist size (a). Cavities with the same bandwidth (16\,kHz here, the limit for $n=1,4$) have the same spatial filtering properties in the large waist limit. Also plotted: variation of the total cavity $g$ factor (b), the roundtrip Gouy phase shift (c) and the mirror radii of curvature (d) of each cavity. As the beam waist varies from $10^{-2}$ to $10^2\,$mm, the cavity geometries (e) go from near-unstable concentric ($\mathcal R \rightarrow L/2$, $g_{1,2} \rightarrow -1$, $\Delta \phi_G \rightarrow 2\pi$) through critically stable confocal ($\mathcal R \rightarrow L$, $g_{1,2} \rightarrow 0$, $\Delta \phi_G \rightarrow \pi$) and up to near-unstable plane-parallel ($\mathcal R \rightarrow +\infty$, $g_{1,2} \rightarrow +1$, $\Delta \phi_G \rightarrow 0$).}
\label{figure:HOM_suppression}
\end{figure}

Having established a lower bound on the cavity bandwidth on the previous section, we determine the maximum level of spatial filtering achievable by the cavity as a function of beam size. This will help in understanding the limitations of the cavity given the constraint imposed on beam size by the expanding atomic cloud. The optical suppression factor of higher-order modes, $S_{nm}$, indicates the fraction of power on the $nm$-th mode with respect to the fundamental mode~\cite{Araya1997}:
\begin{equation}
	S_{nm} = \frac{1}{\sqrt{ 1+\left(\frac{2\finesse}{\pi} \right)^2 \sin^2\left[ \left(n+m\right) \arccos \left( \sqrt{g_1 g_2}~\right) \right] }},
	\label{equation:HOM_suppression}
\end{equation}
where $g_{1,2}$ are the cavity's $g$ factors, $g_i = 1-L/\mathcal R _i$, and $\mathcal R_i$ are the mirrors' radii of curvature. The total cavity $g$ factor, defined as $g_c = g_1g_2$, is a useful quantity to determine if the cavity is geometrically stable ($0 \leq g_c \leq 1$) or otherwise.

We introduce the local Gouy phase for a Hermite-Gaussian beam as~\cite{Siegman1986}
\begin{equation}
	\phi_G = \arctan \left( \frac{z}{z_R}\right),
	\label{equation:local_gouy_phase}
\end{equation}
and the roundtrip Gouy phase shift in the cavity as~\cite{Arai2013}
\begin{equation}
	\Delta \phi_G = 2 \arccos\left(\sqrt{g_1g_2}~\right),
	\label{equation:roundtrip_gouy_phase}
\end{equation}
where $z$ is the position along the optical axis with respect to the center, $z_R = \pi w_0^2 / \lambda$ is the Rayleigh range, and $w_0$ is the beam waist at the center. The Gouy phase shift stems from the transverse spatial confinement of the beam and the consequent spread in transverse momenta~\cite{Feng2001}. Using Eqs.~(\ref{equation:HOM_suppression}),~(\ref{equation:local_gouy_phase}), and~(\ref{equation:roundtrip_gouy_phase}) we can derive an expression for the optical suppression of higher-order modes in terms of the cavity finesse, length, and beam waist:
\begin{equation}
	S_{nm} = \frac{1}{\sqrt{ 1+\left(\frac{2\finesse}{\pi} \right)^2 \sin^2\left[ 2 \left(n+m\right) \arctan \left( \frac{\lambda L}{2\pi w_0^2} \right) \right] }}.
\label{equation:HOM_suppression_waist}
\end{equation}

A lower suppression factor, specially of the first and second order modes ($S_{01}$ and $S_{02}$), indicates that the cavity serves as a better spatial filter, as the circulating field will have a cleaner wavefront. The spatial filtering effect improves the quality of the beam inside the cavity regardless of the origin of the beam distortion, and applies in addition to other means of improving the input beam quality, such as pre-filtering, alignment, and mode matching. Here we do not consider specific input beam properties nor any fluctuations of the cavity parameters. Figure~\ref{figure:HOM_suppression}(a) indicates the relative difference in the intra-cavity build-up of the first and second order spatial modes, which represents the additional improvement in spatial filtering provided by the cavity.

We consider, as an example, three cavities of different length --- 1, 10, and 100\,m --- all having the same bandwidth of $16\,$kHz, which is the lower limit for Bragg diffraction orders $n=1,4$ (Fig.~\ref{figure:HOM_suppression}). In the limit where the waist of the cavity is in the order of interest for atom interferometry, the suppression factors are approximately the same for all cavities having the same bandwidth,
\begin{equation}
	\mathop {\lim}\limits_{\left( \substack{\rm{large} \\ \rm{waist}} \right)} S_{nm} = 
1-\frac{(n+m)^2 c^2 \lambda ^2 }{2 \pi ^4 \Delta \omega_{\rm{min}} ^2 w_0^4} + O\left(\frac{1}{w_0}\right)^8,
\end{equation}
as evidenced by the overlapping curves to the right of Fig.~\ref{figure:HOM_suppression}(a). I.e., for the large beam sizes needed in order to accommodate the thermally expanding clouds, the spatial filtering properties of cavities having the same bandwidth are approximately the same. When the cavity bandwidth is limited for design reasons, the wavefront quality is therefore also limited. Note that despite the fact that the bandwidth limits obtained in the previous section set a very high bar for the finesse of short cavities (e.g., roughly 10000 for $n=1,4$ at $L=1\,$m), they have very similar performance as longer cavities with much smaller finesse (e.g., the 100\,m 100 finesse cavity) for beam waist sizes on the order of a few millimeters. And of course, short cavities with finesses below the limit would have even worse performance in the region of interest than longer cavities operating at the limit.

One always has to mind that the roundtrip Gouy phase shift is not a ratio of $\pi$ so as to avoid bunching of higher-order modes. E.g., $S_{02}$ peaks when $S_{01}$ is minimum indicating confocality for a roundtrip Gouy phase shift of $\Delta \phi_G = \pi$ [Fig.~\ref{figure:HOM_suppression}(a)]; all even modes bunch together at this point. Note that, e.g., for a beam waist size of $w_0 = 5\,\rm{mm}$, the 1 and 10\,m cavities are near-unstable plane-parallel ($\Delta \phi_G \approx 0.006\,\pi$ and $0.063\,\pi$, respectively), while the 100\,m cavity is clearly stable ($\Delta \phi_G \approx 0.587 \pi$).

We distinguish two different limiting factors affecting the maximum allowed beam waist size in the cavity. The first one is the requirement of having a geometrically stable cavity, i.e., having a total $g$ factor of less than what would be experimentally unrealizable. The second one stems from the requirement of achieving a certain level of spatial filtering while staying within the bandwidth bound established in the previous section. We introduce this requirement by constraining the optical suppression factor of the first and second order spatial modes $S_{01}$ and $S_{02}$.

\begin{enumerate}
	\item Geometrical limit:
	\begin{equation}
		g_c \leq g_{\rm{max}}.
	\end{equation}
	\item Optical limit:
	\begin{align}
		S_{01,02} &\leq S_{\rm{max}}, \\
		\Delta \omega &\geq \Delta \omega_{\rm{min}}.
	\end{align}
\end{enumerate}

\begin{figure*}
\centering
\includegraphics[scale=0.9]{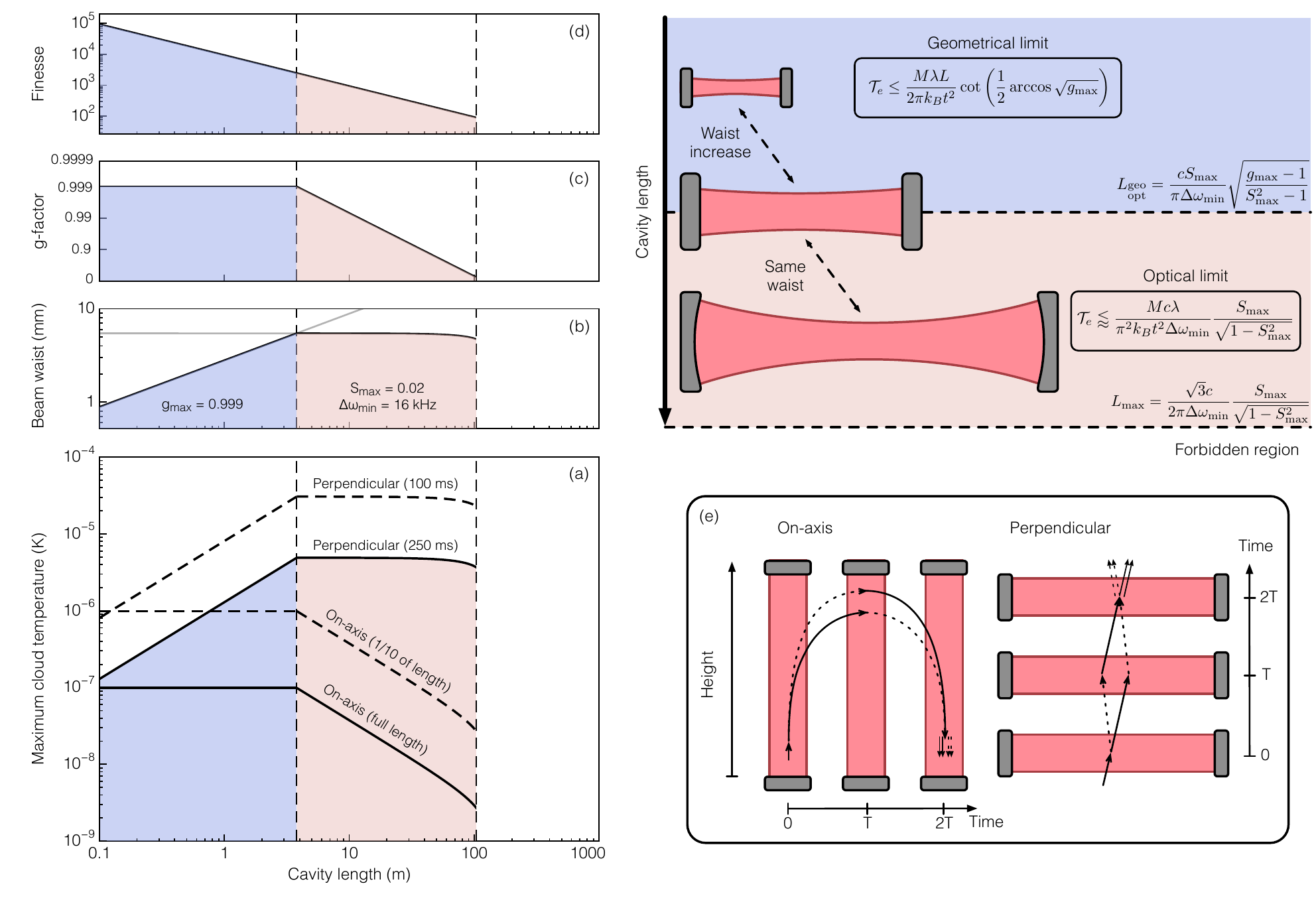}
\caption{Geometrical and optical limits of the cavity-assisted atom interferometer. Beam waist size (b) and cloud temperature limits (a) are derived from a series of constraints. In the geometrical limit the upper bounds are set by the maximum cavity $g$ factor that is experimentally realizable (c). Longer cavities sit more comfortably within geometrical stability but in turn offer worse suppression of higher order spatial modes due to the bandwidth limit. In this region the upper bounds are set by the requirement of achieving a certain level of suppression of the first and second order spatial modes. The maximum cavity finesse is indicated in (d); cavities in the optical limit are by definition at the bandwidth limit, and thus also at the finesse limit. The temperature limits are calculated for two types of interferometric sequences (e). For on-axis sequences we show a case in which the atomic trajectories explore the entire cavity length and one in which they only use 1/10 of the length. For the perpendicular type we show the limits for total measurement times of 100 and 250\,ms.}
\label{figure:cavity_limits}
\end{figure*}

Short cavities will operate in the geometrical limit, as the smaller length comes at the price of putting the cavity very near the edge of geometrical instability. Larger cavities, on the other hand, will be optically limited, while easily maintaining an stable configuration despite the large waist. These upper bounds on beam waist size yield upper bounds on cloud temperature in order to keep the cloud within the confines of the beam (Fig.~\ref{figure:cavity_limits}).

Using Eqs.~(\ref{equation:local_gouy_phase}) and~(\ref{equation:roundtrip_gouy_phase}) we derive an expression for the maximum beam waist given a maximum cavity $g$ factor $g_{\rm{max}}$:
\begin{equation}
	w_{0,\rm{geo}}\left(L\right) = \sqrt{ \frac{L \lambda}{2 \pi } \cot \left( \frac{\arccos \sqrt {g_{\rm{max}}}}{2} \right)}.
\end{equation}
Using Eq.~(\ref{equation:HOM_suppression_waist}) we derive an expression for the maximum beam waist given a maximum suppression factor of the 1st and 2nd order spatial modes, $S_{\rm{max}}$:
\begin{widetext}
\begin{equation}
	w_{0,\rm{opt}}\left(L\right) =  \left( \frac{\lambda ^2}{4 \pi^4 \Delta \omega_{\rm{min}}^2} \frac{2c^2S_{\rm{max}}^2 +\pi^2 L^2 \Delta \omega_{\rm{min}}^2 \left( S_{\rm{max}}^2 - 1 \right) + 2c S_{\rm{max}} \sqrt{c^2S_{\rm{max}}^2 + \pi^2 L^2 \Delta \omega_{\rm{min}}^2 \left( S_{\rm{max}}^2 - 1 \right)} }{1-S_{\rm{max}}^2} \right)^{1/4},
	\label{equation:geo_waist}
\end{equation}

with

\begin{equation}
	L \in \left( 0 , \frac{\sqrt{3} c }{2 \pi  \Delta \omega_{\rm{min}}  } \frac{S_{\rm{max}}}{\sqrt{1-S_{\rm{max}}^2}} \right].
\end{equation}
In the geometrical limit $w_{0,\rm{geo}} < w_{0,\rm{opt}}$, whereas in the optical limit $w_{0,\rm{opt}} < w_{0,\rm{geo}}$.
\end{widetext}

The size of the atomic ensemble after a time $t$ during the experiment is characterized by a Gaussian distribution of width
\begin{equation}
	\sigma_t = \sqrt{\sigma_0^2 + \sigma_v^2 t^2},
	\label{equation:cloud_position_distribution}
\end{equation}
where $\sigma_0$ is the width of the initial position distribution, $\sigma_v = (k_B \mathcal T_e/M)^{1/2}$ is the width of the velocity distribution of temperature $\mathcal T_e$ and mass $M$, and $k_B$ is the Boltzmann constant. Assuming that $\sigma_v t$ is much larger than $\sigma_0$, we can approximate Eq.~(\ref{equation:cloud_position_distribution}) by
\begin{equation}
	\sigma_t \approx \sqrt{\frac{k_B \mathcal T_e}{M}} t.
\end{equation}
The size of the cloud at time $t$ must be, at most, equal to the size of the cavity waist. Thus, the temperature of the atomic ensemble is limited by the maximum waist sizes in either the geometrical or the optical limits:
\begin{equation}
	\mathcal T_e \leq \frac{M w_0^2}{k_B t^2}.
	\label{equation:t_limit}
\end{equation}

A cavity operating in the geometrical limit will have its temperature limited by
\begin{equation}
\mathcal T_e \leq \frac{M\lambda L}{ 2 \pi k_B t^2} \cot \left( \frac{1}{2} \arccos \sqrt{g_{\rm{max}}}\right),
\label{equation:geometrical_limit}
\end{equation}
whereas a cavity operating in the optical limit requires, to first order in $L$:
\begin{equation}
\mathcal T_e \lessapprox \frac{M c \lambda  }{\pi^2  k_B t^2 \Delta \omega_{\rm{min}} } \frac{S_{\rm{max}}}{\sqrt{1-S_{\rm{max}} ^2}}.
\label{equation:optical_limit}
\end{equation}
The approximation given here is valid to first order in $L$. For the exact expression, see Eqs.~(\ref{equation:geo_waist}) and~(\ref{equation:t_limit}).

Lastly, the maximum cavity length allowed under the optical constraints is
\begin{equation}
L_{\rm{max}} = \frac{\sqrt{3} c  }{ 2 \pi \Delta \omega_{\rm{min}} } \frac{S_{\rm{max}}}{\sqrt{1-S_{\rm{max}} ^2}}.
\label{equation:max_length}
\end{equation}
This length limit is independent of the temperature of the atomic ensemble. There are, thus, two factors at play limiting the length of cavities for atom interferometry, and when designing long cavities a sacrifice on either of these limits must be made: either the cavity bandwidth is chosen smaller than $\Delta \omega_{\rm{min}}$, with the consequent problem of increased interaction times and higher power requirements; or the requirements on spatial filtering are relaxed, losing part of the advantage of incorporating the cavity in the first place. 

Owing to how the measurement time scales differently with cavity length for the two types of interferometers considered, the upper bound on cloud temperature scales differently too. In on-axis sequences it scales with 1 in the geometrical limit and $L^{-1}$ in the optical limit, whereas in perpendicular sequences it scales with $L$ and 1 respectively. For the value $g_{\rm{max}}=0.999$ used here, which was determined experimentally to be a safe choice in order to maintain controllability of the cavity~\cite{Wang2017}, an interferometer running on-axis sequences and using the entire cavity length requires sub-$\mu$K temperatures throughout the range. When restricting the atomic trajectories to use only a fraction of the cavity length, these limits are relaxed: if the atoms explore a fraction $1/\alpha$ of the length, the upper bound on cloud temperature scales with $\alpha$. Note that for on-axis interferometers the temperature limits in the optically-limited region are always lower than those in the geometrical limit, independent of the value of $S_{\rm{max}}$.

For running perpendicular sequences the temperature bounds are more forgiving, specially for larger optically-limited cavities. We have presented two cases with total measurement times of 250 and 100\,ms, and the value $S_{\rm{max}}=0.02$ was chosen to obtain substantial suppression of higher order modes. This yields $w_{0,\rm{opt}} \approx 5\,$mm, which is a typical design value~\cite{Riou2017}. Under these constraints large cavities put an upper temperature limit of $4.9\,\mu$K at $L=10\,$m and $3.8\,\mu$K at $L=100\,$m for the $t=250\,$ms case, and $30.9\,\mu$K at $L=10\,$m and $23.8\,\mu$K at $L=100\,$m for the $t=100\,$ms case. For perpendicular sequences cavity stability yields $w_{0,\rm{geo}} \approx 2.8\,$mm at $L=1\,$m, giving upper temperature limits of $ 1.3\,\mu$K and $8.1\,\mu$K for the $t=250\,$ms and $t=100\,$ms cases respectively. However, regardless of $t$, $\mathcal T_e$, or the sequence type, the maximum cavity length is $L_{\rm{max}} \approx 100\,$m, and longer cavities would lie in the forbidden region under these constraints. The higher the order of the diffraction process, the more strict the temperature and length limits are due to the increasing value of $\Delta \omega_{\rm{min}}$. 

These limits are conservative in the sense that the cavity is constrained to accommodate the size of the cloud at the center, where the beam size is smallest. However, both in the geometrical limit and most of the optical limit (when the $g$ factor is close to 1), the cavity is in a near-unstable plane-parallel configuration and thus the size of the beam at the mirrors is approximately equal to the size of the beam waist.

We have assumed that Doppler shifts between atom states are less than the cavity bandwidth. If the cavity is subject to accelerations along the optical axis, the bandwidth must be large enough so as to compensate for the Doppler detuning of the interferometric pulses. The larger bandwidth will result in worse spatial filtering, and therefore in lower upper bounds to cloud temperature in order to maintain the same level of suppression of higher-order modes. To compensate for an increase over the minimum cavity bandwidth, the constraint on the optical suppression of the first and second order spatial modes would have to be relaxed by approximately the same amount. E.g., the situation presented in Fig.~\ref{figure:cavity_limits} is analogous for a cavity with bandwidth 160\,kHz and $S_{\rm{max}}=0.2$.

\section{Summary}

The evolution of an atomic ensemble as it interacts with a pulse of radiation resonating inside an optical cavity has been considered in detail by means of a numerical model. We have shown that there is a lower bound on the cavity bandwidth below which the photon-atom interaction time increases substantially. Cavities with bandwidths below the minimum parametrically push the atomic diffraction process into the long-interaction Bragg and channeling regimes. A cavity operating at the bandwidth limit for the specific diffraction process provides: (1) Maximum power enhancement of the atom optics pulses. (2) Minimum elongation of the interaction time, potentially minimizing the degradation of atomic flux due to velocity selectivity. (3) Best spatial filtering of the interferometric beams.

We have derived the spatial filtering properties of cavities operating at this bandwidth limit as a function of beam waist size. We have further derived beam waist size limits for the interferometer as a function of cavity length, which are divided into what we call the geometrical and optical limits. These limits allow us to determine the maximum temperatures of the atomic ensemble so that the atoms fit within the confines of the beam. A cavity operating with a beam waist size in these limits guarantees that:
(1) the cavity is stable, and
(2) the 1st and 2nd order spatial modes are suppressed below the desired threshold.
In the geometrical limit ---affecting short cavities--- the maximum beam waist size scales with $\sqrt{L}$, whereas in the optical limit --- affecting long cavities --- it stays approximately constant.

A length limit emerges from the optical constraints, restricting the
practicality of long-baseline detectors. This limit is independent of
temperature, scales inversely with the cavity bandwidth and, to first
oder, scales directly with the maximum suppression factor of higher-order modes.

These findings should assist the design of current and future
detectors using two-mirror cavities, and pave the way
towards alternative cavity designs.

\vspace*{1cm}


\begin{acknowledgements}
This work was realized with the financial support of the Defence Science and Technology Laboratory (DSTL) and the UK National Quantum Technology Hub in Sensors and Metrology with EPSRC Grant No. EP/M013294/1.
D.D.B. acknowledges support from the European Commission Horizon 2020 programme under the Q-Sense project Grant No. 691156 (Q-Sense-H2020-MSCA-RISE-2015).
C.M.M.L. acknowledges support from the European Commission Horizon 2020 programme under the Marie Sklodowska-Curie Grant No. 701264.
H.M. is supported by the Ernest Rutherford Fellowship with STFC Grant No. ST/M005844/11.
A.F. is supported by the STFC with Grant No. ST/N000633/1.
M.D.A. would like to thank Nicolas Mielec for helpful discussions about the atom-optics model and Javier \'Alvarez-Vizoso for many useful discussions during the writing of this paper.
\end{acknowledgements}

\subsection*{Author contributions}
M.D.A., C.M.M.L., H.M. and A.F. conceived and designed the study. M.D.A., D.D.B. and A.J. developed the computer model, analyzed and interpreted the data. M.D.A. wrote the paper. D.D.B., A.J., C.M.M.L., H.M. and A.F. provided critical revisions.



\begin{thebibliography}{42}%
\makeatletter
\providecommand \@ifxundefined [1]{%
 \@ifx{#1\undefined}
}%
\providecommand \@ifnum [1]{%
 \ifnum #1\expandafter \@firstoftwo
 \else \expandafter \@secondoftwo
 \fi
}%
\providecommand \@ifx [1]{%
 \ifx #1\expandafter \@firstoftwo
 \else \expandafter \@secondoftwo
 \fi
}%
\providecommand \natexlab [1]{#1}%
\providecommand \enquote  [1]{``#1''}%
\providecommand \bibnamefont  [1]{#1}%
\providecommand \bibfnamefont [1]{#1}%
\providecommand \citenamefont [1]{#1}%
\providecommand \href@noop [0]{\@secondoftwo}%
\providecommand \href [0]{\begingroup \@sanitize@url \@href}%
\providecommand \@href[1]{\@@startlink{#1}\@@href}%
\providecommand \@@href[1]{\endgroup#1\@@endlink}%
\providecommand \@sanitize@url [0]{\catcode `\\12\catcode `\$12\catcode
  `\&12\catcode `\#12\catcode `\^12\catcode `\_12\catcode `\%12\relax}%
\providecommand \@@startlink[1]{}%
\providecommand \@@endlink[0]{}%
\providecommand \url  [0]{\begingroup\@sanitize@url \@url }%
\providecommand \@url [1]{\endgroup\@href {#1}{\urlprefix }}%
\providecommand \urlprefix  [0]{URL }%
\providecommand \Eprint [0]{\href }%
\providecommand \doibase [0]{http://dx.doi.org/}%
\providecommand \selectlanguage [0]{\@gobble}%
\providecommand \bibinfo  [0]{\@secondoftwo}%
\providecommand \bibfield  [0]{\@secondoftwo}%
\providecommand \translation [1]{[#1]}%
\providecommand \BibitemOpen [0]{}%
\providecommand \bibitemStop [0]{}%
\providecommand \bibitemNoStop [0]{.\EOS\space}%
\providecommand \EOS [0]{\spacefactor3000\relax}%
\providecommand \BibitemShut  [1]{\csname bibitem#1\endcsname}%
\let\auto@bib@innerbib\@empty
\bibitem [{\citenamefont {Kasevich}\ and\ \citenamefont
  {Chu}(1991)}]{Kasevich1991}%
  \BibitemOpen
  \bibfield  {author} {\bibinfo {author} {\bibfnamefont {M.}~\bibnamefont
  {Kasevich}}\ and\ \bibinfo {author} {\bibfnamefont {S.}~\bibnamefont {Chu}},\
  }\href {\doibase 10.1103/PhysRevLett.67.181} {\bibfield  {journal} {\bibinfo
  {journal} {Phys. Rev. Lett.}\ }\textbf {\bibinfo {volume} {67}},\ \bibinfo
  {pages} {181} (\bibinfo {year} {1991})}\BibitemShut {NoStop}%
\bibitem [{\citenamefont {Fixler}\ \emph {et~al.}(2007)\citenamefont {Fixler},
  \citenamefont {Foster}, \citenamefont {McGuirk},\ and\ \citenamefont
  {Kasevich}}]{Fixler2007}%
  \BibitemOpen
  \bibfield  {author} {\bibinfo {author} {\bibfnamefont {J.~B.}\ \bibnamefont
  {Fixler}}, \bibinfo {author} {\bibfnamefont {G.~T.}\ \bibnamefont {Foster}},
  \bibinfo {author} {\bibfnamefont {J.~M.}\ \bibnamefont {McGuirk}}, \ and\
  \bibinfo {author} {\bibfnamefont {M.~A.}\ \bibnamefont {Kasevich}},\ }\href
  {\doibase 10.1126/science.1135459} {\bibfield  {journal} {\bibinfo  {journal}
  {Science}\ }\textbf {\bibinfo {volume} {315}},\ \bibinfo {pages} {74}
  (\bibinfo {year} {2007})}\BibitemShut {NoStop}%
\bibitem [{\citenamefont {Rosi}\ \emph {et~al.}(2014)\citenamefont {Rosi},
  \citenamefont {Sorrentino}, \citenamefont {Cacciapuoti}, \citenamefont
  {Prevedelli},\ and\ \citenamefont {Tino}}]{Rosi2014}%
  \BibitemOpen
  \bibfield  {author} {\bibinfo {author} {\bibfnamefont {G.}~\bibnamefont
  {Rosi}}, \bibinfo {author} {\bibfnamefont {F.}~\bibnamefont {Sorrentino}},
  \bibinfo {author} {\bibfnamefont {L.}~\bibnamefont {Cacciapuoti}}, \bibinfo
  {author} {\bibfnamefont {M.}~\bibnamefont {Prevedelli}}, \ and\ \bibinfo
  {author} {\bibfnamefont {G.~M.}\ \bibnamefont {Tino}},\ }\href
  {http://dx.doi.org/doi:10.1038/nature13433} {\bibfield  {journal} {\bibinfo
  {journal} {Nature}\ }\textbf {\bibinfo {volume} {510}},\ \bibinfo {pages}
  {518} (\bibinfo {year} {2014})}\BibitemShut {NoStop}%
\bibitem [{\citenamefont {Weiss}\ \emph {et~al.}(1994)\citenamefont {Weiss},
  \citenamefont {Young},\ and\ \citenamefont {Chu}}]{Weiss1994}%
  \BibitemOpen
  \bibfield  {author} {\bibinfo {author} {\bibfnamefont {D.~S.}\ \bibnamefont
  {Weiss}}, \bibinfo {author} {\bibfnamefont {B.~C.}\ \bibnamefont {Young}}, \
  and\ \bibinfo {author} {\bibfnamefont {S.}~\bibnamefont {Chu}},\ }\href
  {\doibase 10.1007/BF01081393} {\bibfield  {journal} {\bibinfo  {journal}
  {Appl. Phys. B}\ }\textbf {\bibinfo {volume} {59}},\ \bibinfo {pages} {217}
  (\bibinfo {year} {1994})}\BibitemShut {NoStop}%
\bibitem [{\citenamefont {Cadoret}\ \emph {et~al.}(2008)\citenamefont
  {Cadoret}, \citenamefont {de~Mirandes}, \citenamefont {Clad{\'e}},
  \citenamefont {Guellati-Kh{\'e}lifa}, \citenamefont {Schwob}, \citenamefont
  {Nez}, \citenamefont {Julien},\ and\ \citenamefont {Biraben}}]{Cadoret2008}%
  \BibitemOpen
  \bibfield  {author} {\bibinfo {author} {\bibfnamefont {M.}~\bibnamefont
  {Cadoret}}, \bibinfo {author} {\bibfnamefont {E.}~\bibnamefont
  {de~Mirandes}}, \bibinfo {author} {\bibfnamefont {P.}~\bibnamefont
  {Clad{\'e}}}, \bibinfo {author} {\bibfnamefont {S.}~\bibnamefont
  {Guellati-Kh{\'e}lifa}}, \bibinfo {author} {\bibfnamefont {C.}~\bibnamefont
  {Schwob}}, \bibinfo {author} {\bibfnamefont {F.}~\bibnamefont {Nez}},
  \bibinfo {author} {\bibfnamefont {L.}~\bibnamefont {Julien}}, \ and\ \bibinfo
  {author} {\bibfnamefont {F.}~\bibnamefont {Biraben}},\ }\href
  {https://doi.org/10.1103/PhysRevLett.101.230801} {\bibfield  {journal}
  {\bibinfo  {journal} {Phys. Rev. Lett.}\ }\textbf {\bibinfo {volume} {101}},\
  \bibinfo {pages} {230801} (\bibinfo {year} {2008})}\BibitemShut {NoStop}%
\bibitem [{\citenamefont {Bouchendira}\ \emph {et~al.}(2011)\citenamefont
  {Bouchendira}, \citenamefont {Clad{\'e}}, \citenamefont
  {Guellati-Kh{\'e}lifa}, \citenamefont {Nez},\ and\ \citenamefont
  {Biraben}}]{Bouchendira2011}%
  \BibitemOpen
  \bibfield  {author} {\bibinfo {author} {\bibfnamefont {R.}~\bibnamefont
  {Bouchendira}}, \bibinfo {author} {\bibfnamefont {P.}~\bibnamefont
  {Clad{\'e}}}, \bibinfo {author} {\bibfnamefont {S.}~\bibnamefont
  {Guellati-Kh{\'e}lifa}}, \bibinfo {author} {\bibfnamefont {F.}~\bibnamefont
  {Nez}}, \ and\ \bibinfo {author} {\bibfnamefont {F.}~\bibnamefont
  {Biraben}},\ }\href {\doibase 10.1103/PhysRevLett.106.080801} {\bibfield
  {journal} {\bibinfo  {journal} {Phys. Rev. Lett.}\ }\textbf {\bibinfo
  {volume} {106}},\ \bibinfo {pages} {080801} (\bibinfo {year}
  {2011})}\BibitemShut {NoStop}%
\bibitem [{\citenamefont {Schlippert}\ \emph {et~al.}(2014)\citenamefont
  {Schlippert}, \citenamefont {Hartwig}, \citenamefont {Albers}, \citenamefont
  {Richardson}, \citenamefont {Schubert}, \citenamefont {Roura}, \citenamefont
  {Schleich}, \citenamefont {Ertmer},\ and\ \citenamefont
  {Rasel}}]{Schlippert2014}%
  \BibitemOpen
  \bibfield  {author} {\bibinfo {author} {\bibfnamefont {D.}~\bibnamefont
  {Schlippert}}, \bibinfo {author} {\bibfnamefont {J.}~\bibnamefont {Hartwig}},
  \bibinfo {author} {\bibfnamefont {H.}~\bibnamefont {Albers}}, \bibinfo
  {author} {\bibfnamefont {L.~L.}\ \bibnamefont {Richardson}}, \bibinfo
  {author} {\bibfnamefont {C.}~\bibnamefont {Schubert}}, \bibinfo {author}
  {\bibfnamefont {A.}~\bibnamefont {Roura}}, \bibinfo {author} {\bibfnamefont
  {W.~P.}\ \bibnamefont {Schleich}}, \bibinfo {author} {\bibfnamefont
  {W.}~\bibnamefont {Ertmer}}, \ and\ \bibinfo {author} {\bibfnamefont {E.~M.}\
  \bibnamefont {Rasel}},\ }\href {\doibase 10.1103/PhysRevLett.112.203002}
  {\bibfield  {journal} {\bibinfo  {journal} {Phys. Rev. Lett.}\ }\textbf
  {\bibinfo {volume} {112}},\ \bibinfo {pages} {203002} (\bibinfo {year}
  {2014})}\BibitemShut {NoStop}%
\bibitem [{\citenamefont {Tarallo}\ \emph {et~al.}(2014)\citenamefont
  {Tarallo}, \citenamefont {Mazzoni}, \citenamefont {Poli}, \citenamefont
  {Sutyrin}, \citenamefont {Zhang},\ and\ \citenamefont {Tino}}]{Tarallo2014}%
  \BibitemOpen
  \bibfield  {author} {\bibinfo {author} {\bibfnamefont {M.~G.}\ \bibnamefont
  {Tarallo}}, \bibinfo {author} {\bibfnamefont {T.}~\bibnamefont {Mazzoni}},
  \bibinfo {author} {\bibfnamefont {N.}~\bibnamefont {Poli}}, \bibinfo {author}
  {\bibfnamefont {D.~V.}\ \bibnamefont {Sutyrin}}, \bibinfo {author}
  {\bibfnamefont {X.}~\bibnamefont {Zhang}}, \ and\ \bibinfo {author}
  {\bibfnamefont {G.~M.}\ \bibnamefont {Tino}},\ }\href
  {https://doi.org/10.1103/PhysRevLett.113.023005} {\bibfield  {journal}
  {\bibinfo  {journal} {Phys. Rev. Lett.}\ }\textbf {\bibinfo {volume} {113}},\
  \bibinfo {pages} {023005} (\bibinfo {year} {2014})}\BibitemShut {NoStop}%
\bibitem [{\citenamefont {Peters}\ \emph {et~al.}(1999)\citenamefont {Peters},
  \citenamefont {Chung},\ and\ \citenamefont {Chu}}]{Peters1999}%
  \BibitemOpen
  \bibfield  {author} {\bibinfo {author} {\bibfnamefont {A.}~\bibnamefont
  {Peters}}, \bibinfo {author} {\bibfnamefont {K.~Y.}\ \bibnamefont {Chung}}, \
  and\ \bibinfo {author} {\bibfnamefont {S.}~\bibnamefont {Chu}},\ }\href
  {\doibase 10.1038/23655} {\bibfield  {journal} {\bibinfo  {journal} {Nature}\
  }\textbf {\bibinfo {volume} {400}},\ \bibinfo {pages} {849} (\bibinfo {year}
  {1999})}\BibitemShut {NoStop}%
\bibitem [{\citenamefont {Hu}\ \emph {et~al.}(2013)\citenamefont {Hu},
  \citenamefont {Sun}, \citenamefont {Duan}, \citenamefont {Zhou},
  \citenamefont {Chen}, \citenamefont {Zhan}, \citenamefont {Zhang},\ and\
  \citenamefont {Luo}}]{Hu2013}%
  \BibitemOpen
  \bibfield  {author} {\bibinfo {author} {\bibfnamefont {Z.-K.}\ \bibnamefont
  {Hu}}, \bibinfo {author} {\bibfnamefont {B.-L.}\ \bibnamefont {Sun}},
  \bibinfo {author} {\bibfnamefont {X.-C.}\ \bibnamefont {Duan}}, \bibinfo
  {author} {\bibfnamefont {M.-K.}\ \bibnamefont {Zhou}}, \bibinfo {author}
  {\bibfnamefont {L.-L.}\ \bibnamefont {Chen}}, \bibinfo {author}
  {\bibfnamefont {S.}~\bibnamefont {Zhan}}, \bibinfo {author} {\bibfnamefont
  {Q.-Z.}\ \bibnamefont {Zhang}}, \ and\ \bibinfo {author} {\bibfnamefont
  {J.}~\bibnamefont {Luo}},\ }\href {\doibase 10.1103/PhysRevA.88.043610}
  {\bibfield  {journal} {\bibinfo  {journal} {Phys. Rev. A}\ }\textbf {\bibinfo
  {volume} {88}},\ \bibinfo {pages} {043610} (\bibinfo {year}
  {2013})}\BibitemShut {NoStop}%
\bibitem [{\citenamefont {McGuirk}\ \emph {et~al.}(2002)\citenamefont
  {McGuirk}, \citenamefont {Foster}, \citenamefont {Fixler}, \citenamefont
  {Snadden},\ and\ \citenamefont {Kasevich}}]{McGuirk2002}%
  \BibitemOpen
  \bibfield  {author} {\bibinfo {author} {\bibfnamefont {J.~M.}\ \bibnamefont
  {McGuirk}}, \bibinfo {author} {\bibfnamefont {G.~T.}\ \bibnamefont {Foster}},
  \bibinfo {author} {\bibfnamefont {J.~B.}\ \bibnamefont {Fixler}}, \bibinfo
  {author} {\bibfnamefont {M.~J.}\ \bibnamefont {Snadden}}, \ and\ \bibinfo
  {author} {\bibfnamefont {M.~A.}\ \bibnamefont {Kasevich}},\ }\href
  {https://link.aps.org/doi/10.1103/PhysRevA.65.033608} {\bibfield  {journal}
  {\bibinfo  {journal} {Phys. Rev. A}\ }\textbf {\bibinfo {volume} {65}},\
  \bibinfo {pages} {033608} (\bibinfo {year} {2002})}\BibitemShut {NoStop}%
\bibitem [{\citenamefont {Sorrentino}\ \emph {et~al.}(2014)\citenamefont
  {Sorrentino}, \citenamefont {Bodart}, \citenamefont {Cacciapuoti},
  \citenamefont {Lien}, \citenamefont {Prevedelli}, \citenamefont {Rosi},
  \citenamefont {Salvi},\ and\ \citenamefont {Tino}}]{Sorrentino2014}%
  \BibitemOpen
  \bibfield  {author} {\bibinfo {author} {\bibfnamefont {F.}~\bibnamefont
  {Sorrentino}}, \bibinfo {author} {\bibfnamefont {Q.}~\bibnamefont {Bodart}},
  \bibinfo {author} {\bibfnamefont {L.}~\bibnamefont {Cacciapuoti}}, \bibinfo
  {author} {\bibfnamefont {Y.-H.}\ \bibnamefont {Lien}}, \bibinfo {author}
  {\bibfnamefont {M.}~\bibnamefont {Prevedelli}}, \bibinfo {author}
  {\bibfnamefont {G.}~\bibnamefont {Rosi}}, \bibinfo {author} {\bibfnamefont
  {L.}~\bibnamefont {Salvi}}, \ and\ \bibinfo {author} {\bibfnamefont {G.~M.}\
  \bibnamefont {Tino}},\ }\href {\doibase 10.1103/PhysRevA.89.023607}
  {\bibfield  {journal} {\bibinfo  {journal} {Phys. Rev. A}\ }\textbf {\bibinfo
  {volume} {89}},\ \bibinfo {pages} {023607} (\bibinfo {year}
  {2014})}\BibitemShut {NoStop}%
\bibitem [{\citenamefont {Gustavson}\ \emph {et~al.}(1997)\citenamefont
  {Gustavson}, \citenamefont {Bouyer},\ and\ \citenamefont
  {Kasevich}}]{Gustavson1997}%
  \BibitemOpen
  \bibfield  {author} {\bibinfo {author} {\bibfnamefont {T.~L.}\ \bibnamefont
  {Gustavson}}, \bibinfo {author} {\bibfnamefont {P.}~\bibnamefont {Bouyer}}, \
  and\ \bibinfo {author} {\bibfnamefont {M.~A.}\ \bibnamefont {Kasevich}},\
  }\href {\doibase 10.1103/PhysRevLett.78.2046} {\bibfield  {journal} {\bibinfo
   {journal} {Phys. Rev. Lett.}\ }\textbf {\bibinfo {volume} {78}},\ \bibinfo
  {pages} {2046} (\bibinfo {year} {1997})}\BibitemShut {NoStop}%
\bibitem [{\citenamefont {Canuel}\ \emph {et~al.}(2006)\citenamefont {Canuel},
  \citenamefont {Leduc}, \citenamefont {Holleville}, \citenamefont {Gauguet},
  \citenamefont {Fils}, \citenamefont {Virdis}, \citenamefont {Clairon},
  \citenamefont {Dimarcq}, \citenamefont {Borde}, \citenamefont {Landragin},\
  and\ \citenamefont {Bouyer}}]{Canuel2006}%
  \BibitemOpen
  \bibfield  {author} {\bibinfo {author} {\bibfnamefont {B.}~\bibnamefont
  {Canuel}}, \bibinfo {author} {\bibfnamefont {F.}~\bibnamefont {Leduc}},
  \bibinfo {author} {\bibfnamefont {D.}~\bibnamefont {Holleville}}, \bibinfo
  {author} {\bibfnamefont {A.}~\bibnamefont {Gauguet}}, \bibinfo {author}
  {\bibfnamefont {J.}~\bibnamefont {Fils}}, \bibinfo {author} {\bibfnamefont
  {A.}~\bibnamefont {Virdis}}, \bibinfo {author} {\bibfnamefont
  {A.}~\bibnamefont {Clairon}}, \bibinfo {author} {\bibfnamefont
  {N.}~\bibnamefont {Dimarcq}}, \bibinfo {author} {\bibfnamefont {C.~J.}\
  \bibnamefont {Borde}}, \bibinfo {author} {\bibfnamefont {A.}~\bibnamefont
  {Landragin}}, \ and\ \bibinfo {author} {\bibfnamefont {P.}~\bibnamefont
  {Bouyer}},\ }\href {\doibase 10.1103/PhysRevLett.97.010402} {\bibfield
  {journal} {\bibinfo  {journal} {Phys. Rev. Lett.}\ }\textbf {\bibinfo
  {volume} {97}},\ \bibinfo {pages} {010402} (\bibinfo {year}
  {2006})}\BibitemShut {NoStop}%
\bibitem [{\citenamefont {Stockton}\ \emph {et~al.}(2011)\citenamefont
  {Stockton}, \citenamefont {Takase},\ and\ \citenamefont
  {Kasevich}}]{Stockton2011}%
  \BibitemOpen
  \bibfield  {author} {\bibinfo {author} {\bibfnamefont {J.~K.}\ \bibnamefont
  {Stockton}}, \bibinfo {author} {\bibfnamefont {K.}~\bibnamefont {Takase}}, \
  and\ \bibinfo {author} {\bibfnamefont {M.~A.}\ \bibnamefont {Kasevich}},\
  }\href {\doibase 10.1103/PhysRevLett.107.133001} {\bibfield  {journal}
  {\bibinfo  {journal} {Phys. Rev. Lett.}\ }\textbf {\bibinfo {volume} {107}},\
  \bibinfo {pages} {133001} (\bibinfo {year} {2011})}\BibitemShut {NoStop}%
\bibitem [{\citenamefont {Dutta}\ \emph {et~al.}(2016)\citenamefont {Dutta},
  \citenamefont {Savoie}, \citenamefont {Fang}, \citenamefont {Venon},
  \citenamefont {Garrido~Alzar}, \citenamefont {Geiger},\ and\ \citenamefont
  {Landragin}}]{Dutta2016}%
  \BibitemOpen
  \bibfield  {author} {\bibinfo {author} {\bibfnamefont {I.}~\bibnamefont
  {Dutta}}, \bibinfo {author} {\bibfnamefont {D.}~\bibnamefont {Savoie}},
  \bibinfo {author} {\bibfnamefont {B.}~\bibnamefont {Fang}}, \bibinfo {author}
  {\bibfnamefont {B.}~\bibnamefont {Venon}}, \bibinfo {author} {\bibfnamefont
  {C.~L.}\ \bibnamefont {Garrido~Alzar}}, \bibinfo {author} {\bibfnamefont
  {R.}~\bibnamefont {Geiger}}, \ and\ \bibinfo {author} {\bibfnamefont
  {A.}~\bibnamefont {Landragin}},\ }\href
  {https://doi.org/10.1103/PhysRevLett.116.183003} {\bibfield  {journal}
  {\bibinfo  {journal} {Phys. Rev. Lett.}\ }\textbf {\bibinfo {volume} {116}},\
  \bibinfo {pages} {183003} (\bibinfo {year} {2016})}\BibitemShut {NoStop}%
\bibitem [{\citenamefont {Dimopoulos}\ \emph {et~al.}(2007)\citenamefont
  {Dimopoulos}, \citenamefont {Graham}, \citenamefont {Hogan},\ and\
  \citenamefont {Kasevich}}]{Dimopoulos2007}%
  \BibitemOpen
  \bibfield  {author} {\bibinfo {author} {\bibfnamefont {S.}~\bibnamefont
  {Dimopoulos}}, \bibinfo {author} {\bibfnamefont {P.~W.}\ \bibnamefont
  {Graham}}, \bibinfo {author} {\bibfnamefont {J.~M.}\ \bibnamefont {Hogan}}, \
  and\ \bibinfo {author} {\bibfnamefont {M.~A.}\ \bibnamefont {Kasevich}},\
  }\href {\doibase 10.1103/PhysRevLett.98.111102} {\bibfield  {journal}
  {\bibinfo  {journal} {Phys. Rev. Lett.}\ }\textbf {\bibinfo {volume} {98}},\
  \bibinfo {pages} {111102} (\bibinfo {year} {2007})}\BibitemShut {NoStop}%
\bibitem [{\citenamefont {Dimopoulos}\ \emph
  {et~al.}(2008{\natexlab{a}})\citenamefont {Dimopoulos}, \citenamefont
  {Graham}, \citenamefont {Hogan},\ and\ \citenamefont
  {Kasevich}}]{Dimopoulos2008}%
  \BibitemOpen
  \bibfield  {author} {\bibinfo {author} {\bibfnamefont {S.}~\bibnamefont
  {Dimopoulos}}, \bibinfo {author} {\bibfnamefont {P.~W.}\ \bibnamefont
  {Graham}}, \bibinfo {author} {\bibfnamefont {J.~M.}\ \bibnamefont {Hogan}}, \
  and\ \bibinfo {author} {\bibfnamefont {M.~A.}\ \bibnamefont {Kasevich}},\
  }\href {\doibase 10.1103/PhysRevD.78.042003} {\bibfield  {journal} {\bibinfo
  {journal} {Phys. Rev. D}\ }\textbf {\bibinfo {volume} {78}},\ \bibinfo
  {pages} {042003} (\bibinfo {year} {2008}{\natexlab{a}})}\BibitemShut
  {NoStop}%
\bibitem [{\citenamefont {Wolf}\ \emph {et~al.}(2007)\citenamefont {Wolf},
  \citenamefont {Lemonde}, \citenamefont {Lambrecht}, \citenamefont {Bize},
  \citenamefont {Landragin},\ and\ \citenamefont {Clairon}}]{Wolf2007}%
  \BibitemOpen
  \bibfield  {author} {\bibinfo {author} {\bibfnamefont {P.}~\bibnamefont
  {Wolf}}, \bibinfo {author} {\bibfnamefont {P.}~\bibnamefont {Lemonde}},
  \bibinfo {author} {\bibfnamefont {A.}~\bibnamefont {Lambrecht}}, \bibinfo
  {author} {\bibfnamefont {S.}~\bibnamefont {Bize}}, \bibinfo {author}
  {\bibfnamefont {A.}~\bibnamefont {Landragin}}, \ and\ \bibinfo {author}
  {\bibfnamefont {A.}~\bibnamefont {Clairon}},\ }\href
  {https://doi.org/10.1103/PhysRevA.75.063608} {\bibfield  {journal} {\bibinfo
  {journal} {Phys. Rev. A}\ }\textbf {\bibinfo {volume} {75}},\ \bibinfo {eid}
  {063608} (\bibinfo {year} {2007})}\BibitemShut {NoStop}%
\bibitem [{\citenamefont {Kovachy}\ \emph {et~al.}(2015)\citenamefont
  {Kovachy}, \citenamefont {Asenbaum}, \citenamefont {Overstreet},
  \citenamefont {Donnelly}, \citenamefont {Dickerson}, \citenamefont
  {Sugarbaker}, \citenamefont {Hogan},\ and\ \citenamefont
  {Kasevich}}]{Kovachy2015a}%
  \BibitemOpen
  \bibfield  {author} {\bibinfo {author} {\bibfnamefont {T.}~\bibnamefont
  {Kovachy}}, \bibinfo {author} {\bibfnamefont {P.}~\bibnamefont {Asenbaum}},
  \bibinfo {author} {\bibfnamefont {C.}~\bibnamefont {Overstreet}}, \bibinfo
  {author} {\bibfnamefont {C.~A.}\ \bibnamefont {Donnelly}}, \bibinfo {author}
  {\bibfnamefont {S.~M.}\ \bibnamefont {Dickerson}}, \bibinfo {author}
  {\bibfnamefont {A.}~\bibnamefont {Sugarbaker}}, \bibinfo {author}
  {\bibfnamefont {J.~M.}\ \bibnamefont {Hogan}}, \ and\ \bibinfo {author}
  {\bibfnamefont {M.~A.}\ \bibnamefont {Kasevich}},\ }\href
  {https://www.nature.com/articles/doi:10.1038/nature16155} {\bibfield
  {journal} {\bibinfo  {journal} {Nature}\ }\textbf {\bibinfo {volume} {528}},\
  \bibinfo {pages} {530} (\bibinfo {year} {2015})}\BibitemShut {NoStop}%
\bibitem [{\citenamefont {Dimopoulos}\ \emph
  {et~al.}(2008{\natexlab{b}})\citenamefont {Dimopoulos}, \citenamefont
  {Graham}, \citenamefont {Hogan}, \citenamefont {Kasevich},\ and\
  \citenamefont {Rajendran}}]{Dimopoulos2008a}%
  \BibitemOpen
  \bibfield  {author} {\bibinfo {author} {\bibfnamefont {S.}~\bibnamefont
  {Dimopoulos}}, \bibinfo {author} {\bibfnamefont {P.~W.}\ \bibnamefont
  {Graham}}, \bibinfo {author} {\bibfnamefont {J.~M.}\ \bibnamefont {Hogan}},
  \bibinfo {author} {\bibfnamefont {M.~A.}\ \bibnamefont {Kasevich}}, \ and\
  \bibinfo {author} {\bibfnamefont {S.}~\bibnamefont {Rajendran}},\ }\href
  {\doibase 10.1103/PhysRevD.78.122002} {\bibfield  {journal} {\bibinfo
  {journal} {Phys. Rev. D}\ }\textbf {\bibinfo {volume} {78}},\ \bibinfo
  {pages} {122002} (\bibinfo {year} {2008}{\natexlab{b}})}\BibitemShut
  {NoStop}%
\bibitem [{\citenamefont {Graham}\ \emph {et~al.}(2013)\citenamefont {Graham},
  \citenamefont {Hogan}, \citenamefont {Kasevich},\ and\ \citenamefont
  {Rajendran}}]{Graham2013}%
  \BibitemOpen
  \bibfield  {author} {\bibinfo {author} {\bibfnamefont {P.~W.}\ \bibnamefont
  {Graham}}, \bibinfo {author} {\bibfnamefont {J.~M.}\ \bibnamefont {Hogan}},
  \bibinfo {author} {\bibfnamefont {M.~A.}\ \bibnamefont {Kasevich}}, \ and\
  \bibinfo {author} {\bibfnamefont {S.}~\bibnamefont {Rajendran}},\ }\href
  {\doibase 10.1103/PhysRevLett.110.171102} {\bibfield  {journal} {\bibinfo
  {journal} {Phys. Rev. Lett.}\ }\textbf {\bibinfo {volume} {110}},\ \bibinfo
  {pages} {171102} (\bibinfo {year} {2013})}\BibitemShut {NoStop}%
\bibitem [{\citenamefont {Chaibi}\ \emph {et~al.}(2016)\citenamefont {Chaibi},
  \citenamefont {Geiger}, \citenamefont {Canuel}, \citenamefont {Bertoldi},
  \citenamefont {Landragin},\ and\ \citenamefont {Bouyer}}]{Chaibi2016}%
  \BibitemOpen
  \bibfield  {author} {\bibinfo {author} {\bibfnamefont {W.}~\bibnamefont
  {Chaibi}}, \bibinfo {author} {\bibfnamefont {R.}~\bibnamefont {Geiger}},
  \bibinfo {author} {\bibfnamefont {B.}~\bibnamefont {Canuel}}, \bibinfo
  {author} {\bibfnamefont {A.}~\bibnamefont {Bertoldi}}, \bibinfo {author}
  {\bibfnamefont {A.}~\bibnamefont {Landragin}}, \ and\ \bibinfo {author}
  {\bibfnamefont {P.}~\bibnamefont {Bouyer}},\ }\href
  {https://journals.aps.org/prd/abstract/10.1103/PhysRevD.93.021101} {\bibfield
   {journal} {\bibinfo  {journal} {Phys. Rev. D}\ }\textbf {\bibinfo {volume}
  {93}},\ \bibinfo {pages} {021101} (\bibinfo {year} {2016})}\BibitemShut
  {NoStop}%
\bibitem [{\citenamefont {Canuel}\ \emph {et~al.}(2016)\citenamefont {Canuel},
  \citenamefont {Pelisson}, \citenamefont {Amand}, \citenamefont {Bertoldi},
  \citenamefont {Cormier}, \citenamefont {Fang}, \citenamefont {Gaffet},
  \citenamefont {Geiger}, \citenamefont {Harms}, \citenamefont {Holleville},
  \citenamefont {Landragin}, \citenamefont {Lef\`evre}, \citenamefont
  {Lhermite}, \citenamefont {Mielec}, \citenamefont {Prevedelli}, \citenamefont
  {Riou},\ and\ \citenamefont {Bouyer}}]{Canuel2016}%
  \BibitemOpen
  \bibfield  {author} {\bibinfo {author} {\bibfnamefont {B.}~\bibnamefont
  {Canuel}}, \bibinfo {author} {\bibfnamefont {S.}~\bibnamefont {Pelisson}},
  \bibinfo {author} {\bibfnamefont {L.}~\bibnamefont {Amand}}, \bibinfo
  {author} {\bibfnamefont {A.}~\bibnamefont {Bertoldi}}, \bibinfo {author}
  {\bibfnamefont {E.}~\bibnamefont {Cormier}}, \bibinfo {author} {\bibfnamefont
  {B.}~\bibnamefont {Fang}}, \bibinfo {author} {\bibfnamefont {S.}~\bibnamefont
  {Gaffet}}, \bibinfo {author} {\bibfnamefont {R.}~\bibnamefont {Geiger}},
  \bibinfo {author} {\bibfnamefont {J.}~\bibnamefont {Harms}}, \bibinfo
  {author} {\bibfnamefont {D.}~\bibnamefont {Holleville}}, \bibinfo {author}
  {\bibfnamefont {A.}~\bibnamefont {Landragin}}, \bibinfo {author}
  {\bibfnamefont {G.}~\bibnamefont {Lef\`evre}}, \bibinfo {author}
  {\bibfnamefont {J.}~\bibnamefont {Lhermite}}, \bibinfo {author}
  {\bibfnamefont {N.}~\bibnamefont {Mielec}}, \bibinfo {author} {\bibfnamefont
  {M.}~\bibnamefont {Prevedelli}}, \bibinfo {author} {\bibfnamefont
  {I.}~\bibnamefont {Riou}}, \ and\ \bibinfo {author} {\bibfnamefont
  {P.}~\bibnamefont {Bouyer}},\ }\href {http://dx.doi.org/10.1117/12.2228825}
  {\bibfield  {journal} {\bibinfo  {journal} {Proc. SPIE}\
  }\textbf {\bibinfo {volume} {9900}} \bibinfo {pages} {Quantum Optics, 990008} (\bibinfo {year} {2016})}\BibitemShut
  {NoStop}%
\bibitem [{\citenamefont {{Canuel}}\ \emph {et~al.}(2017)\citenamefont
  {{Canuel}}, \citenamefont {{Bertoldi}}, \citenamefont {{Amand}},
  \citenamefont {{Borgo di Pozzo}}, \citenamefont {{Geiger}}, \citenamefont
  {{Gillot}}, \citenamefont {{Henry}}, \citenamefont {{Hinderer}},
  \citenamefont {{Holleville}}, \citenamefont {{Lef{\`e}vre}}, \citenamefont
  {{Merzougui}}, \citenamefont {{Mielec}}, \citenamefont {{Monfret}},
  \citenamefont {{Pelisson}}, \citenamefont {{Prevedelli}}, \citenamefont
  {{Riou}}, \citenamefont {{Rogister}}, \citenamefont {{Rosat}}, \citenamefont
  {{Cormier}}, \citenamefont {{Landragin}}, \citenamefont {{Chaibi}},
  \citenamefont {{Gaffet}},\ and\ \citenamefont {{Bouyer}}}]{Canuel2017}%
  \BibitemOpen
  \bibfield  {author} {\bibinfo {author} {\bibfnamefont {B.}~\bibnamefont
  {{Canuel}}}, \bibinfo {author} {\bibfnamefont {A.}~\bibnamefont
  {{Bertoldi}}}, \bibinfo {author} {\bibfnamefont {L.}~\bibnamefont {{Amand}}},
  \bibinfo {author} {\bibfnamefont {E.}~\bibnamefont {{Borgo di Pozzo}}},
  \bibinfo {author} {\bibfnamefont {R.}~\bibnamefont {{Geiger}}}, \bibinfo
  {author} {\bibfnamefont {J.}~\bibnamefont {{Gillot}}}, \bibinfo {author}
  {\bibfnamefont {S.}~\bibnamefont {{Henry}}}, \bibinfo {author} {\bibfnamefont
  {J.}~\bibnamefont {{Hinderer}}}, \bibinfo {author} {\bibfnamefont
  {D.}~\bibnamefont {{Holleville}}}, \bibinfo {author} {\bibfnamefont
  {G.}~\bibnamefont {{Lef{\`e}vre}}}, \bibinfo {author} {\bibfnamefont
  {M.}~\bibnamefont {{Merzougui}}}, \bibinfo {author} {\bibfnamefont
  {N.}~\bibnamefont {{Mielec}}}, \bibinfo {author} {\bibfnamefont
  {T.}~\bibnamefont {{Monfret}}}, \bibinfo {author} {\bibfnamefont
  {S.}~\bibnamefont {{Pelisson}}}, \bibinfo {author} {\bibfnamefont
  {M.}~\bibnamefont {{Prevedelli}}}, \bibinfo {author} {\bibfnamefont
  {I.}~\bibnamefont {{Riou}}}, \bibinfo {author} {\bibfnamefont
  {Y.}~\bibnamefont {{Rogister}}}, \bibinfo {author} {\bibfnamefont
  {S.}~\bibnamefont {{Rosat}}}, \bibinfo {author} {\bibfnamefont
  {E.}~\bibnamefont {{Cormier}}}, \bibinfo {author} {\bibfnamefont
  {A.}~\bibnamefont {{Landragin}}}, \bibinfo {author} {\bibfnamefont
  {W.}~\bibnamefont {{Chaibi}}}, \bibinfo {author} {\bibfnamefont
  {S.}~\bibnamefont {{Gaffet}}}, \ and\ \bibinfo {author} {\bibfnamefont
  {P.}~\bibnamefont {{Bouyer}}},\ }\href@noop {} {\bibfield  {journal}
  {\bibinfo  {journal} {eprint arXiv:1703.02490}\ } (\bibinfo {year} {2017})},\
  \Eprint {http://arxiv.org/abs/1703.02490} {arXiv:1703.02490
  [physics.atom-ph]} \BibitemShut {NoStop}%
\bibitem [{\citenamefont {Collaboration}(2015)}]{Collaboration2015}%
  \BibitemOpen
  \bibfield  {author} {\bibinfo {author} {\bibfnamefont {L.~S.}\ \bibnamefont
  {Collaboration}},\ }\href {\doibase 10.1088/0264-9381/32/7/074001} {\bibfield
   {journal} {\bibinfo  {journal} {Class. Quantum Grav.}\ }\textbf {\bibinfo
  {volume} {32}},\ \bibinfo {pages} {074001} (\bibinfo {year}
  {2015})}\BibitemShut {NoStop}%
\bibitem [{\citenamefont {McGuirk}\ \emph {et~al.}(2000)\citenamefont
  {McGuirk}, \citenamefont {Snadden},\ and\ \citenamefont
  {Kasevich}}]{McGuirk2000}%
  \BibitemOpen
  \bibfield  {author} {\bibinfo {author} {\bibfnamefont {J.~M.}\ \bibnamefont
  {McGuirk}}, \bibinfo {author} {\bibfnamefont {M.~J.}\ \bibnamefont
  {Snadden}}, \ and\ \bibinfo {author} {\bibfnamefont {M.~A.}\ \bibnamefont
  {Kasevich}},\ }\href {\doibase 10.1103/PhysRevLett.85.4498} {\bibfield
  {journal} {\bibinfo  {journal} {Phys. Rev. Lett.}\ }\textbf {\bibinfo
  {volume} {85}},\ \bibinfo {pages} {4498} (\bibinfo {year}
  {2000})}\BibitemShut {NoStop}%
\bibitem [{\citenamefont {Chiow}\ \emph {et~al.}(2011)\citenamefont {Chiow},
  \citenamefont {Kovachy}, \citenamefont {Chien},\ and\ \citenamefont
  {Kasevich}}]{Chiow2011}%
  \BibitemOpen
  \bibfield  {author} {\bibinfo {author} {\bibfnamefont {S.-w.}\ \bibnamefont
  {Chiow}}, \bibinfo {author} {\bibfnamefont {T.}~\bibnamefont {Kovachy}},
  \bibinfo {author} {\bibfnamefont {H.-C.}\ \bibnamefont {Chien}}, \ and\
  \bibinfo {author} {\bibfnamefont {M.~A.}\ \bibnamefont {Kasevich}},\ }\href
  {\doibase 10.1103/PhysRevLett.107.130403} {\bibfield  {journal} {\bibinfo
  {journal} {Phys. Rev. Lett.}\ }\textbf {\bibinfo {volume} {107}},\ \bibinfo
  {pages} {130403} (\bibinfo {year} {2011})}\BibitemShut {NoStop}%
\bibitem [{\citenamefont {M{\"u}ller}\ \emph
  {et~al.}(2008{\natexlab{a}})\citenamefont {M{\"u}ller}, \citenamefont
  {Chiow}, \citenamefont {Long}, \citenamefont {Herrmann},\ and\ \citenamefont
  {Chu}}]{Mueller2008}%
  \BibitemOpen
  \bibfield  {author} {\bibinfo {author} {\bibfnamefont {H.}~\bibnamefont
  {M{\"u}ller}}, \bibinfo {author} {\bibfnamefont {S.-w.}\ \bibnamefont
  {Chiow}}, \bibinfo {author} {\bibfnamefont {Q.}~\bibnamefont {Long}},
  \bibinfo {author} {\bibfnamefont {S.}~\bibnamefont {Herrmann}}, \ and\
  \bibinfo {author} {\bibfnamefont {S.}~\bibnamefont {Chu}},\ }\href
  {https://doi.org/10.1103/PhysRevLett.100.180405} {\bibfield  {journal}
  {\bibinfo  {journal} {Phys. Rev. Lett.}\ }\textbf {\bibinfo {volume} {100}},\
  \bibinfo {pages} {180405} (\bibinfo {year} {2008}{\natexlab{a}})}\BibitemShut
  {NoStop}%
\bibitem [{\citenamefont {Hamilton}\ \emph
  {et~al.}(2015{\natexlab{a}})\citenamefont {Hamilton}, \citenamefont {Jaffe},
  \citenamefont {Brown}, \citenamefont {Maisenbacher}, \citenamefont {Estey},\
  and\ \citenamefont {M{\"u}ller}}]{Hamilton2015}%
  \BibitemOpen
  \bibfield  {author} {\bibinfo {author} {\bibfnamefont {P.}~\bibnamefont
  {Hamilton}}, \bibinfo {author} {\bibfnamefont {M.}~\bibnamefont {Jaffe}},
  \bibinfo {author} {\bibfnamefont {J.~M.}\ \bibnamefont {Brown}}, \bibinfo
  {author} {\bibfnamefont {L.}~\bibnamefont {Maisenbacher}}, \bibinfo {author}
  {\bibfnamefont {B.}~\bibnamefont {Estey}}, \ and\ \bibinfo {author}
  {\bibfnamefont {H.}~\bibnamefont {M{\"u}ller}},\ }\href
  {https://link.aps.org/doi/10.1103/PhysRevLett.114.100405} {\bibfield
  {journal} {\bibinfo  {journal} {Phys. Rev. Lett.}\ }\textbf {\bibinfo
  {volume} {114}},\ \bibinfo {pages} {100405} (\bibinfo {year}
  {2015}{\natexlab{a}})}\BibitemShut {NoStop}%
\bibitem [{\citenamefont {Hamilton}\ \emph
  {et~al.}(2015{\natexlab{b}})\citenamefont {Hamilton}, \citenamefont {Jaffe},
  \citenamefont {Haslinger}, \citenamefont {Simmons}, \citenamefont {Müller},\
  and\ \citenamefont {Khoury}}]{Hamilton2015a}%
  \BibitemOpen
  \bibfield  {author} {\bibinfo {author} {\bibfnamefont {P.}~\bibnamefont
  {Hamilton}}, \bibinfo {author} {\bibfnamefont {M.}~\bibnamefont {Jaffe}},
  \bibinfo {author} {\bibfnamefont {P.}~\bibnamefont {Haslinger}}, \bibinfo
  {author} {\bibfnamefont {Q.}~\bibnamefont {Simmons}}, \bibinfo {author}
  {\bibfnamefont {H.}~\bibnamefont {Müller}}, \ and\ \bibinfo {author}
  {\bibfnamefont {J.}~\bibnamefont {Khoury}},\ }\href
  {http://science.sciencemag.org/content/349/6250/849} {\bibfield  {journal}
  {\bibinfo  {journal} {Science}\ }\textbf {\bibinfo {volume} {349}},\ \bibinfo
  {pages} {849} (\bibinfo {year} {2015}{\natexlab{b}})}\BibitemShut {NoStop}%
\bibitem [{\citenamefont {Jaffe}\ \emph {et~al.}(2017)\citenamefont {Jaffe},
  \citenamefont {Haslinger}, \citenamefont {Xu}, \citenamefont {Hamilton},
  \citenamefont {Upadhye}, \citenamefont {Elder}, \citenamefont {Khoury},\ and\
  \citenamefont {Müller}}]{Jaffe2017}%
  \BibitemOpen
  \bibfield  {author} {\bibinfo {author} {\bibfnamefont {M.}~\bibnamefont
  {Jaffe}}, \bibinfo {author} {\bibfnamefont {P.}~\bibnamefont {Haslinger}},
  \bibinfo {author} {\bibfnamefont {V.}~\bibnamefont {Xu}}, \bibinfo {author}
  {\bibfnamefont {P.}~\bibnamefont {Hamilton}}, \bibinfo {author}
  {\bibfnamefont {A.}~\bibnamefont {Upadhye}}, \bibinfo {author} {\bibfnamefont
  {B.}~\bibnamefont {Elder}}, \bibinfo {author} {\bibfnamefont
  {J.}~\bibnamefont {Khoury}}, \ and\ \bibinfo {author} {\bibfnamefont
  {H.}~\bibnamefont {Müller}},\ }\href {\doibase 10.1038/nphys4189} {\bibfield
   {journal} {\bibinfo  {journal} {Nat. Phys.}\ }\textbf {\bibinfo {volume}
  {13}},\ \bibinfo {pages} {938} (\bibinfo {year} {2017})}\BibitemShut
  {NoStop}%
\bibitem [{\citenamefont {Haslinger}\ \emph {et~al.}(2017)\citenamefont
  {Haslinger}, \citenamefont {Jaffe}, \citenamefont {Xu}, \citenamefont
  {Schwartz}, \citenamefont {Sonnleitner}, \citenamefont {Ritsch-Marte},
  \citenamefont {Ritsch},\ and\ \citenamefont {Müller}}]{Haslinger2017}%
  \BibitemOpen
  \bibfield  {author} {\bibinfo {author} {\bibfnamefont {P.}~\bibnamefont
  {Haslinger}}, \bibinfo {author} {\bibfnamefont {M.}~\bibnamefont {Jaffe}},
  \bibinfo {author} {\bibfnamefont {V.}~\bibnamefont {Xu}}, \bibinfo {author}
  {\bibfnamefont {O.}~\bibnamefont {Schwartz}}, \bibinfo {author}
  {\bibfnamefont {M.}~\bibnamefont {Sonnleitner}}, \bibinfo {author}
  {\bibfnamefont {M.}~\bibnamefont {Ritsch-Marte}}, \bibinfo {author}
  {\bibfnamefont {H.}~\bibnamefont {Ritsch}}, \ and\ \bibinfo {author}
  {\bibfnamefont {H.}~\bibnamefont {Müller}},\ }\href
  {https://arxiv.org/abs/1704.03577} {\  (\bibinfo {year} {2017})},\ \Eprint
  {http://arxiv.org/abs/1704.03577} {arXiv:1704.03577 [physics.atom-ph]}
  \BibitemShut {NoStop}%
\bibitem [{\citenamefont {Riou}\ \emph {et~al.}(2017)\citenamefont {Riou},
  \citenamefont {Mielec}, \citenamefont {Lef{\`e}vre}, \citenamefont
  {Prevedelli}, \citenamefont {Landragin}, \citenamefont {Bouyer},
  \citenamefont {Bertoldi}, \citenamefont {Geiger},\ and\ \citenamefont
  {Canuel}}]{Riou2017}%
  \BibitemOpen
  \bibfield  {author} {\bibinfo {author} {\bibfnamefont {I.}~\bibnamefont
  {Riou}}, \bibinfo {author} {\bibfnamefont {N.}~\bibnamefont {Mielec}},
  \bibinfo {author} {\bibfnamefont {G.}~\bibnamefont {Lef{\`e}vre}}, \bibinfo
  {author} {\bibfnamefont {M.}~\bibnamefont {Prevedelli}}, \bibinfo {author}
  {\bibfnamefont {A.}~\bibnamefont {Landragin}}, \bibinfo {author}
  {\bibfnamefont {P.}~\bibnamefont {Bouyer}}, \bibinfo {author} {\bibfnamefont
  {A.}~\bibnamefont {Bertoldi}}, \bibinfo {author} {\bibfnamefont
  {R.}~\bibnamefont {Geiger}}, \ and\ \bibinfo {author} {\bibfnamefont
  {B.}~\bibnamefont {Canuel}},\ }\href {\doibase 10.1088/1361-6455/aa7592}
  {\bibfield  {journal} {\bibinfo  {journal} {J. Phys. B}\ }\textbf {\bibinfo
  {volume} {50}},\ \bibinfo {pages} {155002} (\bibinfo {year}
  {2017})}\BibitemShut {NoStop}%
\bibitem [{\citenamefont {Moler}\ \emph {et~al.}(1992)\citenamefont {Moler},
  \citenamefont {Weiss}, \citenamefont {Kasevich},\ and\ \citenamefont
  {Chu}}]{Moler1992}%
  \BibitemOpen
  \bibfield  {author} {\bibinfo {author} {\bibfnamefont {K.}~\bibnamefont
  {Moler}}, \bibinfo {author} {\bibfnamefont {D.~S.}\ \bibnamefont {Weiss}},
  \bibinfo {author} {\bibfnamefont {M.~A.}\ \bibnamefont {Kasevich}}, \ and\
  \bibinfo {author} {\bibfnamefont {S.}~\bibnamefont {Chu}},\ }\href
  {https://link.aps.org/doi/10.1103/PhysRevA.45.342} {\bibfield  {journal}
  {\bibinfo  {journal} {Phys. Rev. A}\ }\textbf {\bibinfo {volume} {45}},\
  \bibinfo {pages} {342} (\bibinfo {year} {1992})}\BibitemShut {NoStop}%
\bibitem [{\citenamefont {M{\"u}ller}\ \emph
  {et~al.}(2008{\natexlab{b}})\citenamefont {M{\"u}ller}, \citenamefont
  {Chiow},\ and\ \citenamefont {Chu}}]{Mueller2008a}%
  \BibitemOpen
  \bibfield  {author} {\bibinfo {author} {\bibfnamefont {H.}~\bibnamefont
  {M{\"u}ller}}, \bibinfo {author} {\bibfnamefont {S.-w.}\ \bibnamefont
  {Chiow}}, \ and\ \bibinfo {author} {\bibfnamefont {S.}~\bibnamefont {Chu}},\
  }\href {https://journals.aps.org/pra/abstract/10.1103/PhysRevA.77.023609}
  {\bibfield  {journal} {\bibinfo  {journal} {Phys. Rev. A}\ }\textbf {\bibinfo
  {volume} {77}},\ \bibinfo {pages} {023609} (\bibinfo {year}
  {2008}{\natexlab{b}})}\BibitemShut {NoStop}%
\bibitem [{\citenamefont {Keller}\ \emph {et~al.}(1999)\citenamefont {Keller},
  \citenamefont {Schmiedmayer}, \citenamefont {Zeilinger}, \citenamefont
  {Nonn}, \citenamefont {D{\"u}rr},\ and\ \citenamefont {Rempe}}]{Keller1999}%
  \BibitemOpen
  \bibfield  {author} {\bibinfo {author} {\bibfnamefont {C.}~\bibnamefont
  {Keller}}, \bibinfo {author} {\bibfnamefont {J.}~\bibnamefont
  {Schmiedmayer}}, \bibinfo {author} {\bibfnamefont {A.}~\bibnamefont
  {Zeilinger}}, \bibinfo {author} {\bibfnamefont {T.}~\bibnamefont {Nonn}},
  \bibinfo {author} {\bibfnamefont {S.}~\bibnamefont {D{\"u}rr}}, \ and\
  \bibinfo {author} {\bibfnamefont {G.}~\bibnamefont {Rempe}},\ }\href
  {\doibase 10.1007/s003400050810} {\bibfield  {journal} {\bibinfo  {journal}
  {Appl. Phys. B}\ }\textbf {\bibinfo {volume} {69}},\ \bibinfo {pages} {303}
  (\bibinfo {year} {1999})}\BibitemShut {NoStop}%
\bibitem [{\citenamefont {Araya}\ \emph {et~al.}(1997)\citenamefont {Araya},
  \citenamefont {Mio}, \citenamefont {Tsubono}, \citenamefont {Suehiro},
  \citenamefont {Telada}, \citenamefont {Ohashi},\ and\ \citenamefont
  {Fujimoto}}]{Araya1997}%
  \BibitemOpen
  \bibfield  {author} {\bibinfo {author} {\bibfnamefont {A.}~\bibnamefont
  {Araya}}, \bibinfo {author} {\bibfnamefont {N.}~\bibnamefont {Mio}}, \bibinfo
  {author} {\bibfnamefont {K.}~\bibnamefont {Tsubono}}, \bibinfo {author}
  {\bibfnamefont {K.}~\bibnamefont {Suehiro}}, \bibinfo {author} {\bibfnamefont
  {S.}~\bibnamefont {Telada}}, \bibinfo {author} {\bibfnamefont
  {M.}~\bibnamefont {Ohashi}}, \ and\ \bibinfo {author} {\bibfnamefont {M.-K.}\
  \bibnamefont {Fujimoto}},\ }\href {\doibase 10.1364/AO.36.001446} {\bibfield
  {journal} {\bibinfo  {journal} {Appl. Opt.}\ }\textbf {\bibinfo {volume}
  {36}},\ \bibinfo {pages} {1446} (\bibinfo {year} {1997})}\BibitemShut
  {NoStop}%
\bibitem [{\citenamefont {Siegman}(1986)}]{Siegman1986}%
  \BibitemOpen
  \bibfield  {author} {\bibinfo {author} {\bibfnamefont {A.~E.}\ \bibnamefont
  {Siegman}},\ }\href@noop {} {\emph {\bibinfo {title} {Lasers}}}\ (\bibinfo
  {publisher} {University Science Books, Mill Valley, California},\ \bibinfo
  {year} {1986})\BibitemShut {NoStop}%
\bibitem [{\citenamefont {Arai}(2013)}]{Arai2013}%
  \BibitemOpen
  \bibfield  {author} {\bibinfo {author} {\bibfnamefont {K.}~\bibnamefont
  {Arai}},\ }\href {https://dcc.ligo.org/LIGO-T1300189/public} {\bibfield
  {journal} {\bibinfo  {journal} {LIGO Document T1300189-v1}\ } (\bibinfo
  {year} {2013})}\BibitemShut {NoStop}%
\bibitem [{\citenamefont {Feng}\ and\ \citenamefont {Winful}(2001)}]{Feng2001}%
  \BibitemOpen
  \bibfield  {author} {\bibinfo {author} {\bibfnamefont {S.}~\bibnamefont
  {Feng}}\ and\ \bibinfo {author} {\bibfnamefont {H.~G.}\ \bibnamefont
  {Winful}},\ }\href {\doibase 10.1364/OL.26.000485} {\bibfield  {journal}
  {\bibinfo  {journal} {Opt. Lett.}\ }\textbf {\bibinfo {volume} {26}},\
  \bibinfo {pages} {485} (\bibinfo {year} {2001})}\BibitemShut {NoStop}%
\bibitem [{\citenamefont {Wang}\ \emph {et~al.}(2017)\citenamefont {Wang},
  \citenamefont {Dovale-\'Alvarez}, \citenamefont {Brown}, \citenamefont
  {Cooper}, \citenamefont {Green}, \citenamefont {T\"oyr\"a}, \citenamefont
  {Miao}, \citenamefont {Mow-Lowry},\ and\ \citenamefont {Freise}}]{Wang2017}%
  \BibitemOpen
  \bibfield  {author} {\bibinfo {author} {\bibfnamefont {H.}~\bibnamefont
  {Wang}}, \bibinfo {author} {\bibfnamefont {M.}~\bibnamefont
  {Dovale-\'Alvarez}}, \bibinfo {author} {\bibfnamefont {D.~D.}\ \bibnamefont
  {Brown}}, \bibinfo {author} {\bibfnamefont {S.}~\bibnamefont {Cooper}},
  \bibinfo {author} {\bibfnamefont {A.}~\bibnamefont {Green}}, \bibinfo
  {author} {\bibfnamefont {D.}~\bibnamefont {T\"oyr\"a}}, \bibinfo {author}
  {\bibfnamefont {H.}~\bibnamefont {Miao}}, \bibinfo {author} {\bibfnamefont
  {C.}~\bibnamefont {Mow-Lowry}}, \ and\ \bibinfo {author} {\bibfnamefont
  {A.}~\bibnamefont {Freise}},\ }\href@noop {} {\bibfield  {journal} {\bibinfo
  {journal} {(in preparation)}\ } }\BibitemShut {NoStop}%
\end{thebibliography}
%


\end{document}